\documentclass[aps,prl,twocolumn,showpacs,10pt,floatfix,superscriptaddress,floatfix, longbibliography]{revtex4-1}
\usepackage{amsmath}
\usepackage{braket}
\usepackage{amssymb}
\usepackage[normalem]{ulem}
\usepackage{ dsfont }
\usepackage{bm}
\usepackage[dvipsnames]{xcolor}
\usepackage{bbm}
\usepackage{mathtools}

\def\ii{\mathrm{i}}
\def\ba#1\ea{\begin{align}#1\end{align}}
\newcommand{\be}{\begin{equation}}
\newcommand{\ee}{\end{equation}}
\newcommand{\f}{\frac}

\begin{document}


\title{Floquet engineered inhomogeneous quantum chaos in critical systems}
\author{Bastien Lapierre}
\email{blapierre@princeton.edu}
\affiliation{Department of Physics, Princeton University, Princeton, New Jersey, 08544, USA}
\author{Tokiro Numasawa}
\affiliation{Institute for Solid State Physics, University of Tokyo, Kashiwa 277-8581, Japan}
\author{Titus Neupert}
\affiliation{Department of Physics, University of Zurich, Winterthurerstrasse 190, 8057 Zurich, Switzerland}
\author{Shinsei Ryu}
\affiliation{Department of Physics, Princeton University, Princeton, New Jersey, 08544, USA}
\date{\today}
\begin{abstract}
We study universal chaotic dynamics of a large class of periodically driven critical systems described by spatially inhomogeneous conformal field theories. By employing an effective curved spacetime approach, we show that the onset of quantum chaotic correlations, captured by the Lyapunov exponent of out-of-time-order correlators (OTOCs), is set by the Hawking temperature of emergent Floquet horizons. Furthermore, scrambling of quantum information is shown to be strongly inhomogeneous, leading to transitions from chaotic to non-chaotic regimes by tuning driving parameters. We finally use our framework to propose a concrete protocol to simulate and measure OTOCs in quantum simulators, by designing an efficient stroboscopic backward time evolution.

\end{abstract}
\maketitle   

\textit{Introduction} ---
Quantum information scrambling characterizes the inaccessibility of globally 
encoded quantum information
to local quantum measurements during a unitary time evolution.
A particularly relevant diagnosis for quantum chaos are out-of-time-order correlators (OTOCs)\cite{Larkin:1969aaa,Hosur:2015ylk,Shenker:2013pqa}, which characterize the growth of local operators and can under certain circumstances be experimentally measured in diverse quantum simulator platforms~\cite{PhysRevA.94.040302, PhysRevLett.124.240505, PhysRevX.11.021010, Li_OTOCNMR_2017, 2017NatPh..13..781G}. 
Such correlators have been shown to decay exponentially at early times with a Lyapunov exponent bounded by the temperature of the system~\cite{Maldacena_2016_chaosbound, PhysRevLett.115.131603, PhysRevLett.123.230606}.
The saturation of the bound has been shown to be achieved for maximally chaotic systems such as the Sachdev-Ye-Kitaev model~\cite{Polchinski_2016} or in the context of holographic conformal field theories (CFTs)~\cite{PhysRevLett.115.131603}.

In parallel, thanks to recent developments in the field of quantum simulation, out-of-equilibrium protocols have become of increasingly experimental relevance to probe new dynamical phases~\cite{Donner,doi:10.1126/science.aal3837, Smith_2019}.
Protocols with periodic driving, so-called Floquet driving, are of particular interest because of their relevance to simulating light-matter coupling.
Certain topological phases can be engineered using using Floquet driving \cite{PhysRevB.79.081406, annurev:/content/journals/10.1146/annurev-conmatphys-031218-013423}, paralleling equilibrium counterparts.
Beyond this  periodically driven systems can realize genuinely non-equilibrium quantum phases, such as time crystals~\cite{PhysRevLett.117.090402, PhysRevLett.116.250401, PhysRevX.7.011026}, or Floquet topological insulators~\cite{PhysRevX.3.031005, PhysRevX.6.021013}. However, addressing questions of ergodicity and quantum chaos in driven many-body systems is generally out of reach.

Recently, a class of exactly solvable driven critical models were introduced~\cite{Wen_2018quench, floquetcftwen, PhysRevResearch.2.023085, lapierre2020geometric, fan_2021}. These consist in gapless field theories with a spatially modulated Hamiltonian density, referred to as inhomogeneous CFTs~\cite{Dubail_2017, 10.21468/SciPostPhys.3.3.019, Gaw_dzki_2018, moosavi2019inhomogeneous}. They have been applied to a variety of non-equilibrium protocols, from periodic to quasi-periodic~\cite{PhysRevResearch.2.033461, PhysRevResearch.3.023044} and random drives~\cite{Wen_2022_random}. 
Besides their intrinsic value in exploring analytically the non-equilibrium dynamics of many-body systems at criticality, these classes of driven systems display rich features, such as spatially localized entanglement~\cite{Fan_2020} and protocols for cooling towards the ground state of a system initially prepared in a thermal state~\cite{wen_Floquet_fridge}.
Furthermore, smooth spatial deformations of gapless critical systems in (1+1)D can be used to engineer curved spacetime geometries such as black hole horizons~\cite{PhysRevResearch.2.023085, Martin_2019, PhysRevResearch.3.L022022}, manifesting as hotspots of energy and sinks of quantum entanglement. These open a pathway for simulations of a black hole atmosphere~\cite{bermond2022anomalous}, as well as black hole creation and evaporation~\cite{non-eq-BH-goto, goto2023scrambling}. 
The non-equilibrium physics of inhomogeneous gapless systems has recently been experimentally realized in quantum gas simulators~\cite{Tajik_2023}, and may be accessible in Rydberg atom platforms, whose tunability in principle allows for highly controllable spatially dependent couplings~\cite{Borish_2020}.

\begin{figure}[t]
	\centering
	\includegraphics[width=1\columnwidth]{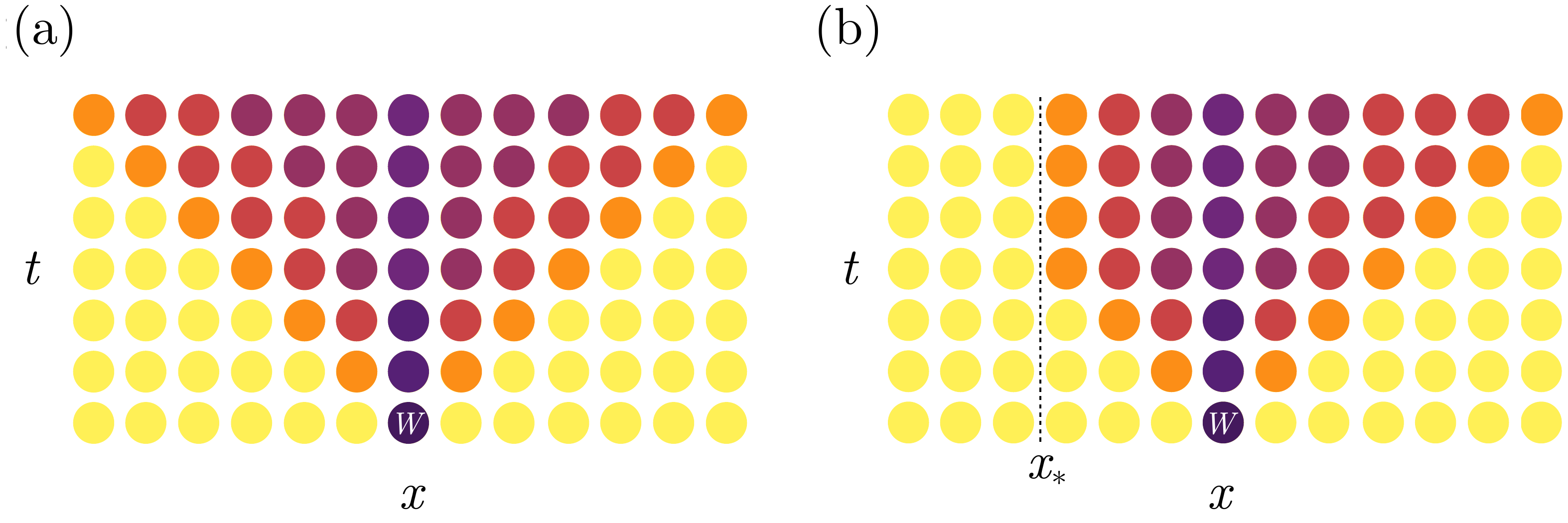}
	\caption{Sketch of spreading of correlations in two different gapless theories. (a) Spreading of correlations for a spatially homogeneous critical system. (b) Stroboscopic spreading of correlations for a periodically driven inhomogeneous CFT in the heating phase. The emergence of a Floquet horizon $x_{*}$ leads to a blockade of correlations through the system at stroboscopic times. As a consequence, a non-trivial OTOC evolution arises only for certain insertions of the operators and not others.}
\label{fig:firstfigOTOC}
\end{figure}

In this Letter, we study the scrambling properties of emergent horizons in a class of periodically driven critical systems.
This class of Floquet drives solely relies on underlying conformal symmetry, and can thus be applied equally to 
free theories, such as Luttinger liquids, 
integrable (rational) CFTs, such as the Ising CFT,
or to strongly correlated theories such as holographic CFTs. 
In the chaotic regime of large central charge, we show that the Floquet horizons lead to an exponential decay of the OTOC. We demonstrate that the Lyapunov exponent satisfies $\lambda_L= 2\pi\Theta_{\text{H}}$, where $\Theta_{\text{H}}$ is the effective Hawking temperature of the horizon. This provides a non-equilibrium and zero temperature analog to the thermal bound on the Lyapunov exponent. Furthermore, we exhibit chaos-to-non-chaos transitions by continuously tuning the driving parameters of the system, as a consequence of the inhomogeneous scrambling illustrated on Fig.~\ref{fig:firstfigOTOC}(b).

One of the main challenges for measuring OTOCs is 
to engineer backward time evolution operators. 
This is one of the reasons why the measurement of OTOCs 
has been so far limited to few-body systems and digital quantum simulators 
\cite{PhysRevA.94.040302, PhysRevLett.124.240505, PhysRevX.11.021010, Li_OTOCNMR_2017, 2017NatPh..13..781G} (
see also other theoretical proposals
\cite{Lantagne_Hurtubise_2020,
yoshida2024universal}).
For many-body quantum systems and quantum field theory, 
however, 
constructing backward time evolution operators poses a significant challenge. 
For example, in quantum field theory with infinite degrees of freedom,
the Hamiltonian is bounded from below but not from above.
Taking this literally poses a fundamental challenge to consider backward time evolution.
In this work, 
we theoretically propose and numerically simulate a simple stroboscopic backward time evolution protocol, as a first step towards a measurement of OTOCs in driven critical quantum systems. 
For rational CFTs,
our Floquet protocol can be used to extract topological data, namely, the braiding phase (the modular $S$ matrix) of primary operators
following earlier works
\cite{Caputa_2016, Gu_2016}.

\textit{Setup} --- We consider a one-dimensional critical model of size $L$, described in the low-energy regime by a (1+1)-dimensional conformal field theory (CFT) of central charge $c$. The system is spatially inhomogeneous, such that it is governed by the Hamiltonian
\begin{equation}
\label{simplebasique}
H[v(x)]=\int_0^L v(x) T_{00}(x)\text{d}x,
\end{equation}
where $v(x)$ is a smooth deformation of the Hamiltonian density, and $T_{00}(x)$ is the stress tensor. For simplicity, we consider a two-step drive between a homogeneous Hamiltonian $H_0$ and an inhomogeneous Hamiltonian $H_1$, for durations $T_0$ and $T_1$. Natural choices for such deformations, called $\mathfrak{sl}(2)$ deformations~\cite{Han_2020}, are of the form $v(x) = \mathfrak{a}+\mathfrak{b}\cos(2\pi x/L)+\mathfrak{c}\sin(2\pi x/L)$, such that they only involve the generators of the \textit{global} conformal algebra. In this case, the stroboscopic evolution of primary fields after $n$ cycles is encoded in a Möbius transformation whose parameters depend on $n$, as well as on $\frac{T_0}{L}$ and $\frac{T_1}{L}$.
The classification of this stroboscopic map dictates the dynamics of correlation functions: for elliptic Möbius transformations, correlation functions oscillate periodically in time, while they grow exponentially for hyperbolic Möbius transformations. These two distinct dynamical behaviours are thus referred to as nonheating and heating phases. While energy grows exponentially in the heating phase, its distribution is strongly peaked only at two spatial locations, and decays exponentially everywhere else. These energy peaks share non-local quantum information, such that the von Neumann entanglement entropy $S_A$ grows linearly in time if the subsystem $A$ (of size $l$) contains one of the horizons, and relaxes down to the ground state entanglement entropy otherwise, $S_A\sim \frac{c}{3}\log(l)$, even when starting from a high-temperature thermal state with initial entropy $S_A\sim \frac{\pi c l}{3\beta}$.

\textit{Emergent Floquet horizons} ---
The stroboscopic description of this class of periodic drives is provided by the Floquet Hamiltonian $H_{\text{F}}$, defined as $U_{\text{F}} = e^{-\ii H_{\text{F}} T}$ for a driving period $T$. For $\mathfrak{sl}(2)$ drives made of an arbitrary number of steps, or for a continuous drive, it takes the general form
\begin{equation}
\label{eq:floquetHam}
H_{\text{F}} = \int_0^L[ v_{\text{eff}}^-(x)T(x)+v_{\text{eff}}^+(x)\overline{T}(x)]\text{d}x,
\end{equation}
where $T_{00}(x)=\frac{1}{2\pi}(T(x)+\overline{T}(x))$, and the effective Floquet velocities for the chiral and anti-chiral parts in the heating phase read~\cite{PhysRevResearch.2.023085}
\begin{equation}
\label{effectiveloreal}
v_{\text{eff}}^{\pm}(x) =2L\Theta_{\text{H}} \frac{\sin[\frac{\pi}{L}(x\mp x_{*,1})]\sin[\frac{\pi}{L}(x\mp x_{*,2})]}{\sin[\frac{\pi}{L}(x_{*,1}-x_{*,2})]},
\end{equation}
where $x_{*,1}$ and $x_{*,2}$ are, respectively, the stable and unstable fixed points of the one-cycle Floquet dynamical map in the heating phase~\footnote{note that $H_{\text{F}}$ bears similarities with the entanglement Hamiltonian of the same Floquet problem, see~\cite{wen_Floquet_fridge}}. We note that in the case of a time-reversal symmetric Floquet drive Eq.~\eqref{effectiveloreal} simplifies to $v_{\text{eff}}^{+}(x)=v_{\text{eff}}^{-}(x)$, such that $H_{\text{F}}$ does not carry any net momentum $\frac{1}{2\pi}(T(x)-\overline{T}(x))$~\cite{SM}. The position $x_{*,1}$ ($L-x_{*,1})$  acts as a sink for chiral (antichiral) quasiparticles. The chiral and antichiral excitations follow lightlike geodesics of the stroboscopic curved spacetimes  $\text{d}s_{\pm}^2=\text{d}x^2-v_{\text{eff}}^{\pm}(x)^2\text{d}t^2$. Expanding the chiral metric $\text{d}s^2_{-}$ around $x_{*,1}$, one finds~\cite{PhysRevResearch.2.023085}
\begin{equation}
\text{d}s^2_{-} = \text{d}x^2-\Theta_{\text{H}}(x-x_{*,1})^2 \text{d}t^2,
\end{equation}
with a real positive constant $\Theta _{\mathrm{H}}$,
and similarly for antichiral quasiparticles around $L-x_{*,1}$. This provides a natural interpretation of $x_{*,1}$ and $L-x_{*,1}$ as  Floquet horizons, carrying a Hawking temperature $\Theta_{\text{H}}$ that sets the energy scale at which energy and entanglement increase,
\begin{equation}
E(t) = \int_0^{L}\langle T_{00}\rangle \text{d}x \sim E_0 e^{\Theta_{\text{H}} n T}, \quad S_A(t)\sim \frac{c}{3}\Theta_{\text{H}} n T.
\end{equation}
We stress that the picture of Floquet horizons generalizes to arbitrary smooth deformations $v(x)$ and number of steps (and continuous drives thereof), with each unstable fixed points carrying a local Hawking temperature characterizing the local energy absorption~\cite{lapierre2020geometric}.

\begin{figure}[t]
	\centering
	\includegraphics[width=\columnwidth]{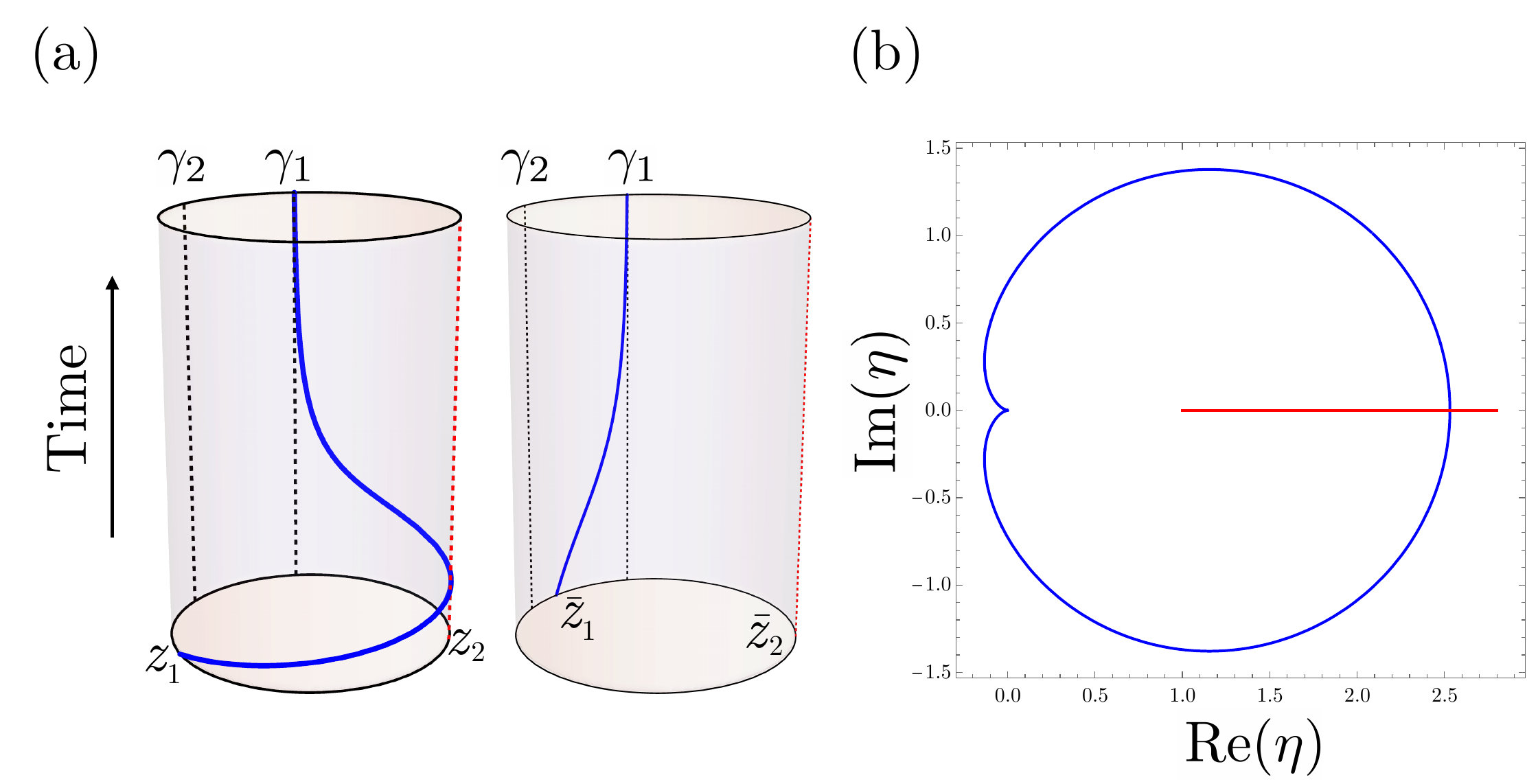}
	\caption{(a) Flow of the primary field $W(z_1,\bar{z}_1)$ under the continuous time evolution with $H_{\text{F}}$ generated by the maps~\eqref{Möbiuseffective2}. \textit{Left:} Flow of the chiral part, that crosses the chiral part of $V(z_2,\bar{z}_2)$ at a time given by~\eqref{eq:timecrossing}. \textit{Right:} Flow of the anti-chiral part, that does not cross $W$. (b) In this case, the cross-ratio $\eta$ crosses the branch cut with the OTOC ordering at the time of the crossing between $W$ and $V$.}
\label{fig:otoc_technical}
\end{figure}

\textit{Inhomogeneous scrambling} ---
Motivated by the fast-scrambling properties of black holes~\cite{Sekino_2008, Shenker_2014_BH}, we study quantum chaotic features of the emergent Floquet horizons in the heating phase. A natural diagnosis of quantum chaos are OTOCs, defined for two local operators $V$ and $W$ as
\begin{equation}
C(x_{1},x_2;t)\sim\langle W(x_1,t)V(x_2)W(x_1,t)V(x_2) \rangle.
\label{fourpointbefore}
\end{equation}
We take $V$ and $W$ to be primary fields of weights $h_V$ and $h_W$, such that the OTOC can be obtained by computing the CFT four-point function
\begin{equation}
\langle\psi_0| W(z_a,\bar{z}_a)W(z_b,\bar{z}_b)V(z_c,\bar{z}_c)V(z_d,\bar{z}_d)| \psi_0\rangle
\end{equation}
which is governed by global conformal invariance by conformal blocks $\mathcal{F}(\eta,\bar{\eta})$, where we define the cross-ratios
\begin{equation}
\eta = \frac{z_{ab}z_{cd}}{z_{ac}z_{bd}},\quad z_{ij} = z_i-z_j,\quad \bar{\eta} = \frac{\bar{z}_{ab}\bar{z}_{cd}}{\bar{z}_{ac}\bar{z}_{bd}},\quad \bar{z}_{ij} = \bar{z}_i-\bar{z}_j.
\end{equation}
The OTOC evolution for homogeneous CFTs at equilibrium is found by complexifying time $t_i\mapsto t_i+\ii \epsilon_i$ and using the time ordering $\epsilon_a>\epsilon_c>\epsilon_b>\epsilon_d$. A crossing of the branch cut $[1,\infty)$ of $\mathcal{F}(\eta,\bar{\eta})$ at early times from either $\eta$ or $\bar{\eta}$ results in an early time exponential growth given by the Lyapunov exponent $\lambda_L=2\pi\beta^{-1¨}$~~\cite{Maldacena_2016_chaosbound, PhysRevLett.115.131603, PhysRevLett.123.230606}.
We now compute the OTOC from a quantum quench with the Floquet Hamiltonian~\eqref{effectiveloreal} in the heating phase~\footnote{Treating the problem as a single quantum quench circumvents  ambiguities on the crossing of the branch cut due to evolving the system with only stroboscopic time steps}. The conformal map encoding the time evolution of the fields $V$ and $W$ with $H_{\text{F}}$ for a  time $t$ is~\cite{SM}
\begin{equation}
\label{Möbiuseffective2}
\tilde{z}_i(t) = \frac{\alpha z_i+\beta}{\beta^* z_i+\alpha^*},\quad \tilde{\bar{z}}_i(t) = \frac{\alpha \bar{z}_i+\beta}{\beta^* \bar{z}_i+\alpha^*},
\end{equation}
with $\alpha$, $\beta$ given by
\begin{equation}
\label{coefMöbiusnontrviv}
\begin{split}
\alpha &=\cosh[\pi\Theta_{\text{H}}t]+\frac{\gamma_1+\gamma_2}{\gamma_1-\gamma_2}\sinh[\pi\Theta_{\text{H}} t], \\
\beta &= -\frac{2\gamma_1\gamma_2}{\gamma_1-\gamma_2}\sinh[\pi\Theta_{\text{H}} t],
\end{split}
\end{equation}
where $\gamma_i=e^{2\pi\ii x_{*,i}/L}$.
Thus the late-time asymptotics are
$\lim_{t\rightarrow\infty} \tilde{z}_i(t)=\lim_{t\rightarrow\infty} \tilde{\bar{z}}_i(t) =\gamma_1$.
This property ensures that only one cross-ratio $\eta$ or $\bar{\eta}$ crosses the branch cut and not the other [see Fig.~\ref{fig:otoc_technical}(a)], leading to nontrivial OTOC evolution. 
The condition for $\eta$ to cross the branch cut, as illustrated on Fig.~\ref{fig:otoc_technical}(b), reads
\begin{equation}
\label{eq:timecrossing}
\tanh(\pi\Theta_{\text{H}} t) = \frac{(z_2-z_1)(\gamma_1-\gamma_2)}{(z_1+z_2)(\gamma_1+\gamma_2)-2(z_1z_2+\gamma_1\gamma_2)},
\end{equation}
and a similar condition can be derived for $\bar{\eta}$ to cross the branch cut. Let us now assume that $\eta$ crosses the branch cut. The late-time asymptotics read
\begin{equation}
    \eta\sim e^{-2\pi \Theta_{\text{H}} t}\epsilon_{12}\epsilon_{34}\frac{\sin[\frac{\pi}{L}(x_1-x_{*,2})]\sin[\frac{\pi}{L}(x_2-x_{*,1})]}{\sin[\frac{\pi}{L}(x_1-x_{*,1})]\sin[\frac{\pi}{L}(x_2-x_{*,2})]},
\label{asympasymp2}
\end{equation}
with $\epsilon_{ij}=\epsilon_i-\epsilon_j$.
In the limit of large central charge, with $h_W/c$ fixed and small and $h_V$ fixed and large, $\mathcal{F}(\eta)$ can be approximated following~\cite{Fitzpatrick_2014_CFT}
\begin{equation}
\mathcal{F}(\eta)=\frac{1}{\left(1-\frac{24\pi\ii h_W}{c\eta}\right)^{h_V}}.
\label{resultOTOCsl2}
\end{equation}
This subsequently leads to an exponential growth of the OTOC, characteristic of the maximally chaotic regime of holographic CFTs.
The lack of translational invariance of the OTOC inherited from the inhomogeneous Hamiltonian~\eqref{simplebasique} is manifest in \eqref{asympasymp2}. This leads to an inhomogenous Butterfly velocity $v_{B}(x)$~\cite{Das_2022}, which equals the Floquet velocity,
\begin{equation}
v_B(x) = v_{\text{eff}}^-(x),
\end{equation}
and similarly with $v_{\text{eff}}^+(x)$ if $\bar{\eta}$ crosses the branch cut instead. The butterfly velocity goes to zero at the Floquet horizons, prohibiting correlations to leak through the horizons at stroboscopic times. In the heating phase the Lyapunov exponent satisfies the relation
\begin{equation}
\lambda_L = 2\pi\Theta_{\text{H}},
\end{equation}
which is an out-of-equilibrium analog to the thermal bound on quantum chaos, $\lambda_L = 2\pi\beta^{-1}$. We stress that while the system is initialized in the ground state of the uniform CFT, the Floquet horizons of the heating phase provide an emergent local temperature. The system appears as if it was thermal with respect to that temperature for OTOCs measured near the respective horizon. 
Furthermore, an inhomogeneous nontrivial OTOC evolution with polynomial decay arises for driven compactified free boson CFTs  with irrational compactification radius~\cite{Caputa_2017, Kudler_Flam_2020_chaosdiag}, realized in, e.g., the $XXZ$ model for irrational values of the anisotropy. 
We finally note that the above discussion generalizes to Floquet drives with arbitrary inhomogeneous Hamiltonians~\cite{SM}, beyond $\mathfrak{sl}(2)$ algebra.

\textit{Rindler drives and chaos transitions} --- We now consider a CFT on the real line, and define a Floquet drive between a homogeneous Hamiltonian $H_0$ and a Rindler Hamiltonian~\cite{PhysRevB.93.235119} $H_1=\int_{-\infty}^{\infty} h|x|T_{00}(x)\text{d}x$. 
The time evolution generated by $H_0$ and $H_1$ can be expressed in the high-frequency limit as the coordinate transformations~\cite{SM}
\ba
x^+(x,t) &=  x e^{h\tau t} + \f{2 (1-\tau)}{h \tau} e^{\f{h}{2}\tau t}\sinh \f{h\tau t}{2},  \notag \\
x^-(x,t) &=  x e^{-h\tau t} - \f{2 (1-\tau)}{h \tau} e^{-\f{h}{2}\tau t}\sinh \f{h\tau t}{2} . \label{eq:RPfloquetcoord}
\ea
with $\tau=T_1/(T_0+T_1)$, and $x^{\mp}$ the coordinates for left and right movers respectively.
Thus, a single Floquet horizon appears at
$x_* = \frac{1}{h}(1-1/\tau)$, which acts as a stable fixed point for the holomorphic sector, and as an unstable fixed point for the anti-holomorphic sector. By repeating the previous calculation of the OTOC, we conclude that non-trivial OTOC time evolution only happens if the initial operator insertion positions $x_1$ and $x_2$ of $W$ and $V$ are on the same side of the Floquet horizon $x_*$, as illustrated on Fig.~\ref{fig:firstfigOTOC}(b). As one continuously tunes $T_0$ and $T_1$, the Floquet horizon $x_*$ shifts position on the real line. For fixed operator initial positions $x_1$ and $x_2$, this leads to transitions from chaotic to non-chaotic regimes dynamically induced at a fixed position.

\begin{figure}[t]
\begin{center}
	\includegraphics[width=0.85\columnwidth]{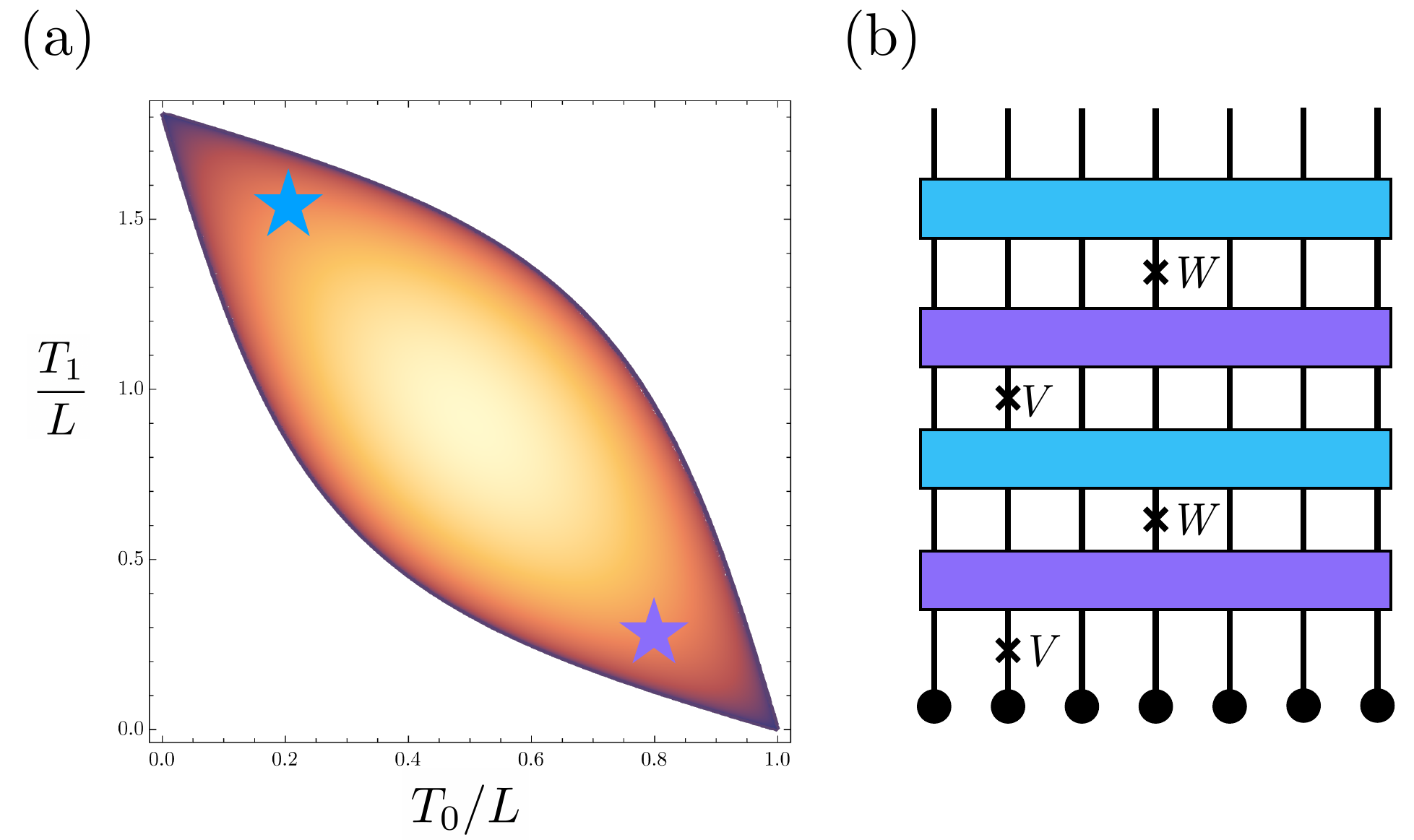}
	\caption{\label{fig:otoc_backward} (a) Heating region of the phase diagram for a two-step $\mathfrak{sl}(2)$ Floquet drive. The colors scale represents the heating rate, or Hawking temperature $\Theta_{\text{H}}$. The two stars represent two sets of driving parameters $\left(\frac{T_0}{L},\frac{T_1}{L}\right)$ and $\left(\frac{\widetilde{T}_0}{L},\frac{\widetilde{T}_1}{L}\right)$ that satisfy~\eqref{reversalcondition}. (b) The OTOC can be measured by combining the two driving sequences in order to stroboscopically evolve forward and backward in time, which just amounts to modulating the durations $T_0$ and $T_1$ without changing the driving Hamiltonians.  }
\end{center}
\end{figure}

\textit{Backward time evolution} --- 
As a first step towards a measurement of the inhomogeneous scrambling of the OTOC on quantum simulators, we propose a simple and concrete protocol that provides a stroboscopic backward time evolution of driven inhomogeneous gapless chains.
The backward time evolution is designed by switching the sign of the \textit{unphysical} Floquet Hamiltonian~\eqref{eq:floquetHam}. A natural way to achieve this for spatially inhomogeneous drives is to switch the driving parameters $(T_0,T_1)$ to $(\widetilde{T}_0,\widetilde{T}_1)$ after driving the system for $n$ cycles, and switch the order of the unitaries $U_0$ and $U_1$, as illustrated on Fig.~\ref{fig:otoc_backward}(b).
The parameters $(\widetilde{T}_0,\widetilde{T}_1)$ are defined such that only the stable and unstable fixed points swap, $x_{*,1}\xleftrightarrow{}x_{*,2}$,  leading to
\begin{equation}
\label{reversalcondition}
\widetilde{U}_{\text{F}}=e^{-\ii\widetilde{H}_F(\widetilde{T}_0+\widetilde{T}_1)}=e^{\ii H_{\text{F}}(T_0+T_1)}= U_{\text{F}}^{\dagger},
\end{equation}
as can be seen from the explicit expression of $H_{\text{F}}$ in~\eqref{effectiveloreal}~\footnote{Similar backward evolution protocols were discussed in the context of electromagnetic fields in periodically driven cavities~\cite{Martin_2019}.}. We stress that this time-reversing protocol applies to \textit{any} two-step drives with $v_i(x)>0$ for $i=0,1$~\cite{SM}, as well as to any initial state, be it a pure or thermal state. In the heating phase this leads to a Floquet engineered ``evaporation'' of the emergent Floquet horizons~\cite{non-eq-BH-goto}, with the half-system entanglement entropy linearly decreasing back to the initial state entanglement, and energy density relaxing back to the initial state.  Experimentally, the advantage of such a procedure is that there is no need to switch the sign of the coupling of the driving Hamiltonians, and the backward evolution is implemented directly by changing the duration of each driving Hamiltonian.

\textit{Numerical calculations} --- At the field theory level our time-reversal procedure leads to a perfect heat erasure and initial state retrieval, even after an arbitrary number of driving cycles in the heating phase. However, lattice effects due to excitation of higher-energy modes are expected to affect the fidelity of the time-reversal procedure, as non-linearities in the spectrum become relevant. We numerically check the robustness of this procedure on a driven fermionic lattice, with inhomogeneous Hamiltonians of the form
\begin{equation}
H[v(j)] = -\frac{1}{2}\sum_{j=1}^{L-1}v(j)c_j^{\dagger}c_{j+1}+h.c..
\label{lattice_hamitlonians_deformed}
\end{equation}
\begin{figure}[t]
\begin{center}
	\includegraphics[width=\columnwidth]{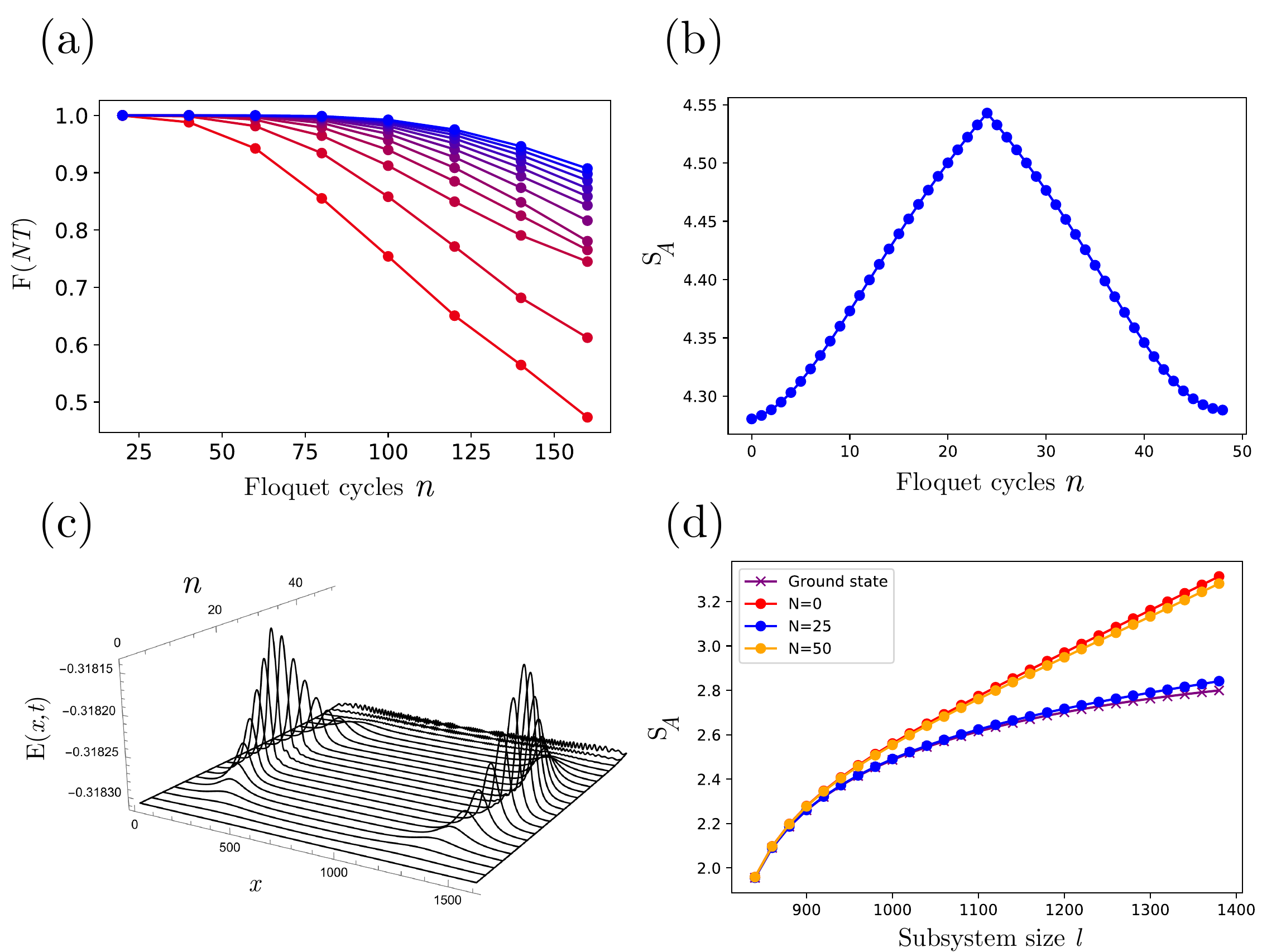}
	\caption{\label{fig:figure_numerics_paper} Numerical results on a free fermion chain driven in the heating phase, and relaxed back to the initial state using the time-reversal protocol. (a) Fidelity $F(n)$ of the time-reversal procedure as function of $N$ for different system sizes $L=100, 200, ..., 1200$ (red to blue). (b) Starting from a Gibbs state at temperature $\beta^{-1}=0.002$ and $L=2000$, the half-system entanglement entropy $S_A(t)$ grows linearly, and decays linearly during the time-reversal procedure, reaching back the initial thermal entanglement entropy. (c) We observe the emergence of Floquet horizons in the energy density evolution, as well as their evaporation. (d) Scaling of the entanglement entropy on the subsystem size $l$. At initial and final times, $S_A$ follows the volume law of a thermal state, while it follows the ground state scaling after 25 cycles deep in the heating phase. }
\end{center}
\end{figure}
The dynamics of such a driven chain at half-filling is well-described by $c=1$ free boson CFT in both heating and nonheating phases for large stroboscopic times as well as pure and thermal initial states~\cite{thermaleffects}. As a first check, we study the fidelity of the time-reversal procedure, defined as
\begin{equation}
F(n)= |\langle \psi_0|\widetilde{U}_{\text{F}}^{n}U_{\text{F}}^{n}|\psi_0\rangle|^2.
\end{equation}
As $n$ increases, lattice effects become more dominant and eventually the CFT time-reversal procedure breaks down. Nevertheless, we find that the fidelity stays close to one when starting from the ground state, even for more than a hundred Floquet cycles, as shown on Fig.~\ref{fig:figure_numerics_paper}(c). Similarly, the entanglement entropy $S_A(t)$ is shown on Fig.~\ref{fig:figure_numerics_paper}(b) to initially grow linearly with the first driving sequence and decrease linearly with the second one, reminiscent of the Page curve evolution of entanglement found in moving mirror CFTs~\cite{PhysRevLett.126.061604}. The two driving sequences operate as creation and evaporation of the emergent Floquet horizons as can be seen in the energy density evolution on Fig.~\ref{fig:figure_numerics_paper}(a).
We finally remark that our time-reversal procedure provides a natural way to interpolate continuously between volume law and ground state scaling of entanglement when the initial state is a thermal state at temperature $\beta^{-1}\gg l$ (such that $S_A\sim \frac{\pi l}{3\beta}$ at equilibrium) and the entanglement cut $A$ does not contain any Floquet horizons, as shown on Fig.~\ref{fig:figure_numerics_paper}(d). Such operation is a full cycle of the ``Floquet's refrigerator'' introduced in~\cite{wen_Floquet_fridge}, as it additionally allows to transit from area to volume law.

\textit{Conclusions} --- 
Our Floquet protocol could be implemented in quantum simulators such as Rydberg atom arrays which can be tuned to conformal critical points 
\cite{Fendley_2004,
PhysRevA.86.041601,
bernien2017, Keesling_2019,
rader2019floating,
Slagle_2021}. In particular, it would provide a promising route to experimentally measure
not only
the chaotic behaviors of CFTs, but also
the universal topological properties of some rational CFTs. These include the braiding properties (the modular $S$ matrix) of, for instance, the Ising CFT \cite{Caputa_2016, Gu_2016}.
Turning to theory, it would be desirable to understand further quantum-information-theoretic properties of the emergent Floquet horizons, such as information metrics; first steps in this directions were initiated in~\cite{deboer2023quantum}. Moreover, it would be interesting to uncover the bridge between driven CFTs and moving mirrors. Such setups display a similar phenomenology~\cite{Martin_2019}, and can also simulate black hole evaporation and the Page curve of entanglement entropy~\cite{PhysRevLett.126.061604, Akal_2021}.

\begin{acknowledgments}\textit{Acknowledgments} --- B.L. thanks R. Chitra for inspiring discussions. B.L. acknowledges funding from 
the Swiss National Foundation for Science (Postdoc.Mobility Grant No. 214461). T.N.\ acknowledges support from the Swiss National Science Foundation through a Consolidator Grant (iTQC, TMCG-$2\_$213805).  
S.R.~is supported
by a Simons Investigator Grant from
the Simons Foundation (Award No.~566116).
This work is supported by
the Gordon and Betty Moore Foundation through Grant
GBMF8685 toward the Princeton theory program. 

\end{acknowledgments}

\bibliographystyle{apsrev4-1}
\bibliography{ref}

\clearpage
\begin{widetext}
.
\setcounter{equation}{0}
\renewcommand\theequation{S\arabic{equation}}
\section*{Supplementary Material}

\subsection{Effective description of $\mathfrak{sl}(2)$ drives}
\label{app:effectivedescri}

The aim of this section is to provide a derivation of the Möbius transformation~\eqref{coefMöbiusnontrviv} encoding the continuous time evolution after a quantum quench with the Floquet Hamiltonian~\eqref{eq:floquetHam}, and to draw a simple quasiparticle picture for the emergence of the Floquet horizons after such a quantum quench. In all generality, an $\mathfrak{sl}(2)$ Hamiltonian takes the form 
\begin{equation}
H = \int_0^L [v(x)T(x)+\bar{v}(x)\overline{T}(x)]\text{d}x,
\end{equation}
with chiral and anti-chiral velocities given by
\begin{equation}
\label{eq:generalSL2deformation}
v(x) = \sigma_0 + \sigma_+ \cos\frac{2\pi x}{L}+\sigma_-\sin\frac{2\pi x}{L}, \quad \bar{v}(x) = \bar{\sigma}_0 + \bar{\sigma}_+ \cos\frac{2\pi x}{L}+\bar{\sigma}_-\sin\frac{2\pi x}{L}.
\end{equation}
The Möbius transformation encoding the continuous time evolution with such inhomogeneous Hamiltonian takes the form~\cite{Han_2020}
\begin{equation}
\tilde{z}(t) = \frac{\alpha z+\beta}{\beta^* z+\alpha^*},\quad \tilde{\bar{z}}(t) = \frac{\bar{\alpha} \bar{z}+\bar{\beta}}{\bar{\beta}^* \bar{z}+\bar{\alpha}^*},
\end{equation}
with coefficients given by 
\begin{equation}
\alpha = \cosh\left(\frac{\pi \mathcal{C} t}{L}\right)+\frac{\ii \sigma_0}{\mathcal{ C}}\sinh\left(\frac{\pi \mathcal{C} t}{L}\right),\quad \quad \beta =\frac{\ii }{\mathcal{ C}}(\sigma_++\ii \sigma_-)\sinh\left(\frac{\pi \mathcal{C} t}{L}\right),
\end{equation}
\begin{equation}
\bar{\alpha} = \cosh\left(\frac{\pi \mathcal{C} t}{L}\right)+\frac{\ii \bar{\sigma}_0}{\mathcal{ C}}\sinh\left(\frac{\pi \mathcal{C} t}{L}\right),\quad \quad \bar{\beta} =\frac{\ii }{\mathcal{ C}}(\bar{\sigma}_+-\ii \bar{\sigma}_-)\sinh\left(\frac{\pi \mathcal{C} t}{L}\right),
\end{equation}
with $\mathcal{C}=\sqrt{|-\sigma_0^2+\sigma_+^2+\sigma_-^2|}$.
Note in particular that the coefficients $\beta$ and $\bar{\beta}$ are in general distinct if the quenching Hamiltonian has the same deformation in both sectors, i.e., $v(x)=\bar{v}(x)$. Let us now focus on the case of the $\mathfrak{sl}(2)$ Floquet Hamiltonian $H_{\text{F}}$ given by~\eqref{eq:floquetHam} in the heating phase. In this case, one readily finds that
\begin{align}
\begin{split}
\sigma_0=-\ii L\Theta_{\text{H}} \frac{\gamma_1+\gamma_2}{\gamma_1-\gamma_2}, \quad\quad &\bar{\sigma}_0=\sigma_0,\\
\sigma_+ = \ii L\Theta_{\text{H}}  \frac{1+\gamma_1\gamma_2}{\gamma_1-\gamma_2}, \quad\quad &\bar{\sigma}_+ = \sigma_+,\\
\sigma_-= L\Theta_{\text{H}} \frac{\gamma_1\gamma_2-1}{\gamma_1-\gamma_2}, \quad\quad  &\bar{\sigma}_-= -\sigma_-.
\end{split}
\end{align}
This implies $\alpha=\bar{\alpha}$ and $\beta=\bar{\beta}$ in this case, because of the peculiar asymmetry between $v_{\text{eff}}^+(x)$ and $v_{\text{eff}}^-(x)$ in~\eqref{effectiveloreal}, which correctly leads to the transformation~\eqref{coefMöbiusnontrviv}. In the case of time-reversal-symmetric drives, $\gamma_1=\gamma_2^*$, i.e., $x_*=x_{*,1}=L-x_{*,2}$, one finds
\begin{equation}
\label{timerevereffectivel}
v_{\text{eff}}(x)=v_{\text{eff}}^+(x)=v_{\text{eff}}^-(x) = 2L\Theta_{\text{H}} \frac{\sin[\frac{\pi}{L}(x-x_{*})]\sin[\frac{\pi}{L}(x+x_{*})]}{\sin[\frac{2\pi x_{*}}{L}]},
\end{equation}
i.e., the deformations of the chiral and anti-antichiral sectors are identical. Furthermore, in the case of time-reversal symmetric drives $\sigma_-=\bar{\sigma}_-=0$, which again leads to $\alpha=\bar{\alpha}$ and $\beta=\bar{\beta}$.
\begin{figure}[h]
	\centering
	\includegraphics[width=0.7\columnwidth]{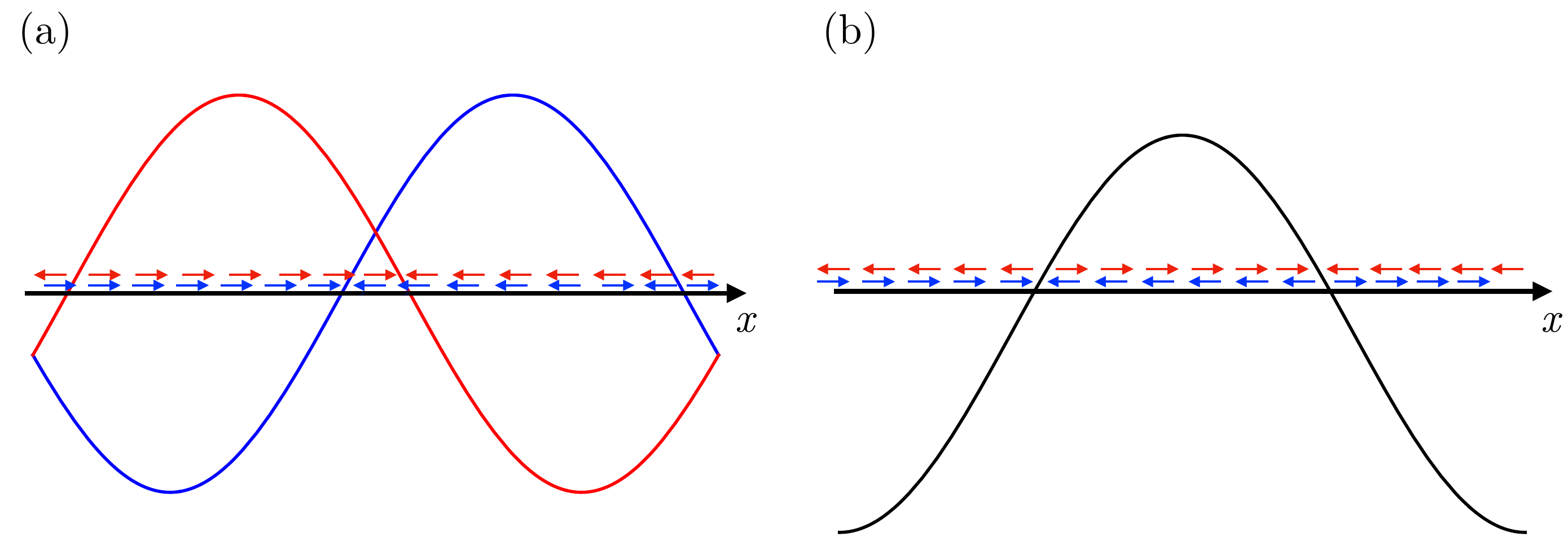}
	\caption{Quasiparticle picture for a quench with the $\mathfrak{sl}(2)$ Floquet Hamiltonian. (a) The Floquet Hamiltonian $H_{\text{F}}$ for a non-time-reversal symmetric drive leads to different components $v_{\text{eff}}^-(x)$ and $v_{\text{eff}}^+(x)$ for chiral and anti-chiral sectors, see~\eqref{eq:floquetHam}, as illustrated with the blue and red curves. In this case the Floquet velocities for both sectors are distinct, leading to two distinct fixed points for each chirality: one acts as a quasiparticle source and one as a sink. This leads to the emergence of two Floquet horizons at symmetric positions. (b) The Floquet Hamiltonian $H_{\text{F}}$ for a time-reversal symmetric drive. In this case both Floquet velocities coincide, and are given by~\eqref{timerevereffectivel}, such that the unstable fixed point of the chiral sector coincides with the stable fixed point of the antichiral sector and vice-versa.}
\label{fig:time_rev_effective}
\end{figure}

Although non-time-reversal symmetric drives and time-reversal symmetric drive have a different effective description, we now present a simple quasiparticle picture that unifies both cases from a Floquet Hamiltonian standpoint, as illustrated on Fig.~\ref{fig:time_rev_effective}. In the non-symmetric case, two distinct fixed points emerge for each sector: a stable and an unstable fixed point, both characterized by their local heating rate. At the unstable fixed point, quasiparticles are getting repelled as the sign of the deformation changes (quasiparticles of the same chirality but at different positions $x_1$ and $x_2$, such that $v_+(x_1)<0<v_+(x_2)$, propagate in opposite directions), while they get attracted at the stable fixed point, leading to a chiral emergent horizon. The same mechanism leads to an anti-chiral horizon at a symmetric position, see Fig.~\ref{fig:time_rev_effective}(a). In the time-reversal symmetric case, the unstable fixed point of the chiral sector merges with the stable fixed point of the anti-chiral sector and vice-versa as shown on Fig.~\ref{fig:time_rev_effective}(b), which again leads to two emergent Floquet horizons. An important distinction between these two effective descriptions of the Floquet drive is that the Floquet Hamiltonian in the non-symmetric case carries a non-zero momentum, i.e., $T-\overline{T}\neq 0$, leading to asymmetric quasiparticle propagation, as illustrated on Fig.~\ref{fig:twopointfunctions}(a). 
\begin{figure}[h]
	\centering
	\includegraphics[width=0.65\columnwidth]{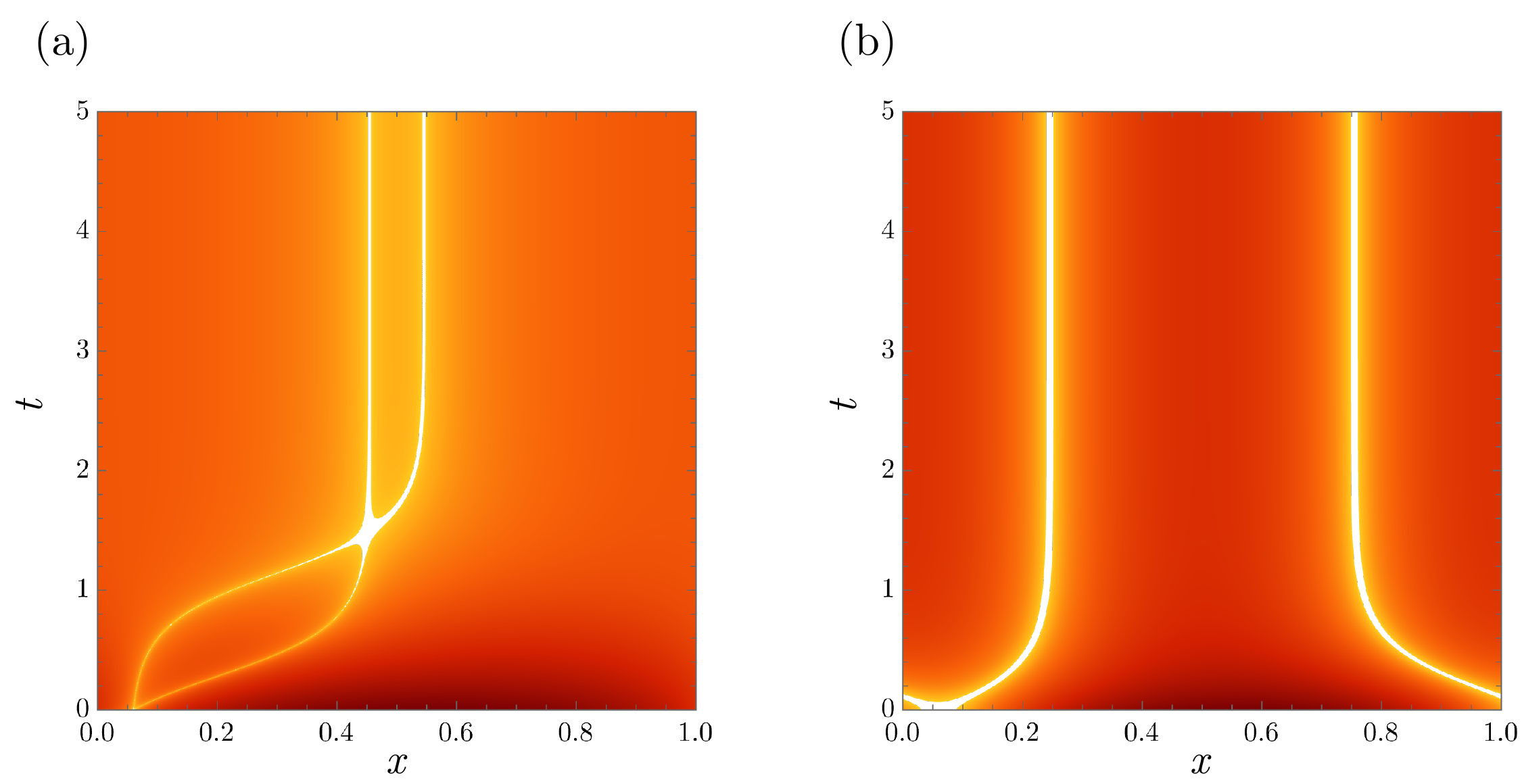}
	\caption{Dynamical two-point function $|\langle \Phi(x,t)\Phi(x_0,0)\rangle|$ for an arbitrary primary field $\Phi$ of weight $(h,\bar{h})$, periodic boundaries, after a quantum quench with the Floquet Hamiltonian $H_{\text{F}}$ for (a) a non-time-reversal symmetric driving sequence and (b) a time-reversal symmetric driving sequence, corresponding respectively to the Hamiltonian deformations from Fig.~\ref{fig:time_rev_effective}(a) and Fig.~\ref{fig:time_rev_effective}(b) (the system size has been set to unity). The non-zero momentum of the Floquet Hamiltonian in the non-symmetric driving sequence leads to chiral and anti-chiral quasiparticles propagating in the same direction, even with periodic boundary conditions.  }
\label{fig:twopointfunctions}
\end{figure}
In this case the effective velocity profiles can have a different sign between the two chiralities, leading to a propagation of chiral and anti-chiral quasiparticles in the same direction, despite the boundaries being periodic and the two sectors thus being uncoupled. This  emergent phenomena of the Floquet drive is however forbidden for time-reversal symmetric drives (see Fig.~\ref{fig:twopointfunctions}(b)), and is a direct consequence of the zero net momentum carried by the Floquet Hamiltonian in this case. It will be interesting to further study from a holographic standpoint the consequences on the AdS bulk of the finite momentum carried by the Floquet Hamiltonian in the non-symmetric case~\cite{jiang2024new}.

\subsection{OTOCS for inhomogeneous Floquet CFTs}
\label{App:OTOCcomputation}

Let us derive the OTOC evolution under a generic $\mathfrak{sl}(2)$ periodic driving.
Our arguments rely on the knowledge of the form of the Floquet Hamiltonian~\eqref{effectiveloreal}, which is independent of the details of the drive, e.g., the numbers of driving steps in one cycle, and applies to both discrete step drives as well as continuous drives. 
We consider a quantum quench with $H_{\text{F}}$, starting from the conformal vacuum $|0\rangle$. We compute the time evolution of the four point function
\begin{equation}
\frac{1}{\mathcal{N}}\langle 0|e^{\ii H_{\text{F}} t}W(z_a,\bar{z}_a)e^{-\ii H_{\text{F}} t}e^{\ii H_{\text{F}} t}W(z_b,\bar{z}_b)e^{-\ii H_{\text{F}} t}V(z_c,\bar{z}_c)V(z_d,\bar{z}_d)|0 \rangle,
\label{app:fourpointfunction}
\end{equation}
where $V$ and $W$ are primary fields of weights $(h_V,\bar{h}_V)$, $(h_W,\bar{h}_W)$, and $\mathcal{N}$ a normalization factor of the form $\langle VV\rangle \langle WW \rangle$. The time evolution of primary fields with $H_{\text{F}}$ in Heisenberg picture is given by the continuous time Möbius transformation~\eqref{Möbiuseffective2}. Thus, the four point function can be expressed as 
\begin{equation}
\frac{1}{\mathcal{N}}\left(\frac{\partial \tilde{z}_a}{\partial z}\frac{\partial \tilde{z}_b}{\partial z}\right)^{h_W}\left(\frac{\partial \tilde{\bar{z}}_a}{\partial \bar{z}}\frac{\partial \tilde{\bar{z}}_b}{\partial \bar{z}}\right)^{\bar{h}_W}\langle 0|W(\tilde{z}_a,\tilde{\bar{z}}_a)W(\tilde{z}_b,\tilde{\bar{z}}_b)V(z_c,\bar{z}_c)V(z_d,\bar{z}_d)|0 \rangle.
\label{fourpointafter}
\end{equation}
We note that the derivative factors in front of the four-point function will not contribute to the time evolution of the OTOC, as they are cancelled by the normalization $\mathcal{N}$.
The OTO ordering is obtained by using the correct $\ii\epsilon$ prescription, where we introduce the complexified time $\tau_j=\ii t_j + \epsilon_j$. Such prescription is given by the ordering of limits $\epsilon_a>\epsilon_c>\epsilon_b>\epsilon_d\rightarrow0$.
From global conformal invariance, it is known that the four point function~\eqref{app:fourpointfunction} is governed by $\mathcal{F}(\eta,\bar{\eta})$, where we defined the cross-ratio
\begin{equation}
\eta = \frac{z_{ab}z_{cd}}{z_{ac}z_{bd}},\quad z_{ij} = z_i-z_j\quad \bar{\eta} = \frac{\bar{z}_{ab}\bar{z}_{cd}}{\bar{z}_{ac}\bar{z}_{bd}},\quad \bar{z}_{ij} = \bar{z}_i-\bar{z}_j.
\end{equation}
The non-trivial behaviour of the OTOC is obtained as the cross-ratio crosses such branch cut to go on the second Riemann sheet~\cite{Roberts_2015}. On a thermal CFT of infinite size at equilibrium, $\eta$ crosses the real line at
\begin{equation}
    \frac{\epsilon_{ab}\epsilon_{cd}}{\epsilon_{ac}\epsilon_{bd}},
\end{equation}
which for the OTO prescription is larger than one, thus crossing the branch cut, while $\bar{\eta}$ does not cross the branch cut. We choose to restrict to pure initial states, as considering the OTOC computation from a thermal initial at finite system size $L$ is beyond the scope of this work. We note that the approach from~\cite{Das_2022} has an ambiguity due to the fact that the time is only evaluated at discrete stroboscopic values, such that it makes it difficult to judge whether or not the cross-ratio $\eta$ or $\bar{\eta}$ has crossed the real axis in between two stroboscopic times $n(T_0+T_1)$ and $(n+1)(T_0+T_1)$, and the analytic continuation to a complexified of a discrete stroboscopic time is unclear. This motivates the use of the Floquet Hamiltonian $H_{\text{F}}$,
such that time is continuous and the crossing of the real axis can explicitly be observed. The final result for the OTOC will then only be valid for stroboscopic times in order to agree with the OTOC for a Floquet drive.

We can now explicitly compute the time evolution of the cross-ratios $\eta$ and $\bar{\eta}$ after the quantum quench with $H_{\text{F}}$, with the correct $\ii\epsilon$ prescription of the OTOC. 
The cross-ratio is
\begin{equation}
\eta = -\frac{64z_1z_2\gamma_1\gamma_2\sin(\frac{\pi}{L}(x_1-x_{*,1}))\sin(\frac{\pi}{L}(x_1-x_{*,2}))\sin(\frac{\pi}{L}(x_2-x_{*,1}))\sin(\frac{\pi}{L}(x_2-x_{*,2}))\sin(\pi\Theta_{\text{H}}\epsilon_{13})\sin(\pi\Theta_{\text{H}}\epsilon_{24})}{[(z_1-z_2)(\gamma_1-\gamma_2)\cosh(\pi\Theta_{\text{H}}(t+\ii\epsilon_{13}))+[(z_1+z_2)(\gamma_1+\gamma_2)-2(\gamma_1\gamma_2+z_1z_2)]\sinh(\pi\Theta_{\text{H}}(t+\ii\epsilon_{13}))][\epsilon_{13}\rightarrow\epsilon_{24}]},
\end{equation}
introducting the notation $\epsilon_{ij} = \epsilon_i - \epsilon_j$.
Note that the anti-holomorphic cross-ratio $\bar{\eta}$ can similarly be computed.
In order for the cross-ratio to be $\mathcal{O}(1)$, and cross the branch cut, we arrive to the condition
\begin{equation}
\tanh(\pi\Theta_{\text{H}} t) = \frac{(z_2-z_1)(\gamma_1-\gamma_2)}{(z_1+z_2)(\gamma_1+\gamma_2)-2(z_1z_2+\gamma_1\gamma_2)}.
\end{equation}
 Conversely the condition for $\bar{\eta}$ to cross the branch cut reads
\begin{equation}
\tanh(\pi\Theta_{\text{H}} t) = \frac{(\bar{z}_2-\bar{z}_1)(\gamma^*_1-\gamma^*_2)}{(\bar{z}_1+\bar{z}_2)(\gamma^*_1+\gamma^*_2)-2(\bar{z}_1\bar{z}_2+\gamma^*_1\gamma^*_2)}.
\end{equation}
Let us now assume that $\eta$ crosses the branch cut but not $\bar{\eta}$. The large time asymptotics thus read
\begin{equation}
    \eta\sim e^{-2\pi \Theta_{\text{H}} t}2\pi^2\Theta_{\text{H}}^2\epsilon_{12}\epsilon_{34}\frac{\sin\frac{\pi}{L}(x_1-x_{*,2})\sin\frac{\pi}{L}(x_2-x_{*,1})}{\sin\frac{\pi}{L}(x_1-x_{*,1})\sin\frac{\pi}{L}(x_2-x_{*,2})} = e^{-2\pi \Theta_{\text{H}} t}2\pi^2\Theta_{\text{H}}^2\epsilon_{12}\epsilon_{34}v_*(x_1,x_2).
\label{asympasymp}
\end{equation}
From~\eqref{asympasymp} it is clear that the  Lyapunov exponent is $\lambda_L = 2\pi \Theta_{\text{H}}$. The inhomogeneous butterfly velocity $v_B(x)$ reads~\cite{Das_2022}
\begin{align}
v_B(x)  &\equiv\lambda_L v_*(x_1,x_2)\left(\frac{\partial v_{*}(x_1,x_2)}{\partial x_2}\right)^{-1}\\\nonumber&= 2L\Theta_{\text{H}}\frac{\sin\frac{\pi}{L}(x-x_{*,1})\sin\frac{\pi}{L}(x-x_{*,2})}{\sin\frac{\pi}{L}(x_{*,1}-x_{*,2})}.
\end{align}
Thus, the butterfly velocity associated to the crossing of the holomorphic cross-ratio equals the holomorphic part of the effective velocity for the Floquet drive~\eqref{effectiveloreal},
\begin{equation}
v_B(x) = v_{\text{eff}}^{+}(x).
\end{equation}
For completeness, let us now assume that $\bar{\eta}$ crosses the branch cut, such that the long time asymptotics read (upon replacement $x_{*,1}\mapsto L-x_{*,2}$, $x_{*,2}\mapsto L-x_{*,1}$)
\begin{equation}
\bar{\eta} = \sim e^{-2\pi \Theta_{\text{H}} t}2\pi^2\Theta_{\text{H}}^2\epsilon_{12}\epsilon_{34}\frac{\sin\frac{\pi}{L}(x_1+x_{*,1})\sin\frac{\pi}{L}(x_2+x_{*,2})}{\sin\frac{\pi}{L}(x_1+x_{*,2})\sin\frac{\pi}{L}(x_2+x_{*,1})}.
\end{equation}
Thus it becomes clear that in this case,
\begin{equation}
v_B(x) = v_{\text{eff}}^{-}(x).
\end{equation}
We thus conclude that the Butterfly velocity is equal to the chiral or antichiral Floquet velocity, depending on whether $\eta$ or $\bar{\eta}$ crosses the branch cut after the quantum quench with $H_{\text{F}}$. In other words, the inhomogeneous scrambling of quantum information is carried by the chiral or the antichiral part of the Floquet Hamiltonian depending on the initial positions of the fields $V$ and $W$.

The generalization of the $\mathfrak{sl}(2)$ Floquet drive to any smooth velocity profiles $v(x)>0$ was considered in~\cite{lapierre2020geometric, fan_2021}. In this case, the deformed Hamiltonians do not only involve the global conformal algebra spanned by $\{L_0,L_1,L_{-1}\}$, but the full Virasoro algebra. While the Virasoro algebra is infinite dimensional, a similar strategy still holds to solve the Floquet dynamics: one can write down a one-cycle diffeomorphism $f_{\pm}$ ($\pm$ stands for the chiral and anti-chiral components respectively) whose $n$-th iteration encodes the stroboscopic evolution of any (quasi-)primary field. For simplicity let us consider a two-step drive between a homogenous Hamiltonian $H_0$ and an arbitrary deformed Hamiltonian $H_1$ with velocity $v_1(x)$. Such one-cycle transformation takes the form
\begin{equation}
\label{onecyclediffeotucoco}
f_{\pm} = f_1^{-1}(f_1(x\mp T_0) \mp  v_1T_1),
\end{equation}
where the diffeomorphism associated to the velocity profile $v_1(x)$ of $H_1$ reads
\begin{equation}
f_1(x) = \int_0^x \text{d}x'\frac{v_1}{v_1(x')}, \quad \frac{1}{v_1}=\frac{1}{L}\int_{0}^L\frac{\text{d}x}{v_1(x)}.
\end{equation}
The heating phase is diagnosed by fixed points of $f_{\pm}$, or higher periodic points, i.e., fixed points of the $m$-th iteration of $f_{\pm}$ for any $m$, denoted by $f_{\pm}^m$. The number of such fixed points can be arbitrary, leading to many emergent Floquet horizons at positions $x_{*,m}^{\mp}$, each characterized by a local Hawking temperature, or heating rate, $\frac{1}{m(T_0+T_1)}\log f'^m_{\pm}(x_{*,m}^{\mp})$.
While it is tempting to generalize the reasoning detailed in this section to this general class of drives, we stress that our approach based on the Floquet Hamiltonian hardly applies in this case. The reason is that deriving a Floquet Hamiltonian in the general case is a complicated task as the Virasoro algebra is infinite dimensional. On the other hand, dealing with iterations of the diffeomorphisms $f_{\pm}$ does not allow for an analytic continuation to a complexified time, as time $n$ only appears through $n$-th composition of  $f_{\pm}$, i.e., there is no closed form for the $n$-th iteration of a general diffeomorphism of the circle. Nonetheless it is clear that the fixed point picture for OTOC evolution will also hold in such a general case, and is illustrated in Fig.~\ref{otoc_general_config}. If the operator $V$ is inserted at time $t=0$ close to a Floquet horizon at position $x^+_{*,i}$, the Floquet velocity will scale as
\begin{equation}
v_{\text{eff}}(x)\sim  \frac{|\log(f_-'(x_{*,i}^{+}))|}{T} (x -  x_{*,i}^{+}),
\end{equation}
where $T$ is the period of the drive, and $f'_{-}$ is the one-cycle diffeomorphism.
We thus conclude that the Butterfly velocity will vanish close to the emergent fixed points.
From the above considerations it is clear that the decay of the OTOC will be governed by the local Hawking temperature of the horizon $x_{*,i}^{+}$, given by $\Theta_{H,i}=\frac{|\log(f'_{-}(x^+_{*,i}))|}{T}$.
\begin{figure}[h!]
\begin{center}
	\includegraphics[width=120mm]{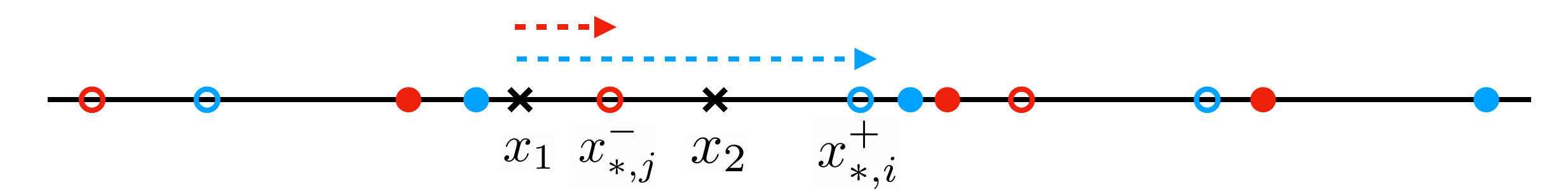}
	\caption{General geometric picture for the operator evolution under a Floquet drive made of arbitrary steps and arbitrary velocity profiles $v_i(x)>0$. The stable (unstable) fixed points of the chiral sector are shown as blue hollow (filled) circles, while they are illustrated in red for the anti-chiral sector. The blue (red) dashed arrow illustrates the flow generated by the $n$-th composition of 1-cycle diffeomorphism $f_{-}$ ($f_{+}$). Note that the alternating structure between stable and unstable fixed points within a given sector is a general property of 1-cycle diffeomorphisms~\cite{lapierre2020geometric}. Here, a non-trivial winding of $\eta$ is shown as $x_1$ goes through to $x_2$ when flowing to the stable fixed point $x_{*,i}^+$. On the other hand, $\bar{\eta}$ will not cross the branch cut as the anti-chiral sector flows to the stable fixed point $x_{*,j}^-$ in between $x_1$ and $x_2$. The obtained OTOC will be governed by the local Hawking temperature associated to the fixed point $x^+_{*,i}$. }
\label{otoc_general_config}
\end{center}
\end{figure}

\subsection{Rindler-Floquet drives}

In this section, we give details on the Floquet dynamics generated by a two-step drive between the uniform CFT Hamiltonian and the Rindler Hamiltonian on an infinite system.
\subsubsection{Floquet dynamics with the Rindler Hamiltonian}
Let us consider the Floquet evolution between the uniform Hamiltonian $H_0$ and the Rindler Hamiltonian $H_1$,
\be
\ket{\psi(t)} = (e^{-\ii T_0 H_0  }e^{-\ii T_1 H_1 })^n \ket{\psi_0}.
\ee
In the high-frequency limit, the dynamics reduces to the evolution by the Hamiltonian
\be
H_{\text{eff}} = a  L_0 + \f{1}{2}(a-ib)L_1 + \f{1}{2}(a+ib)L_{-1}  + a  \bar{L}_0 + \f{1}{2}(a+ib)\bar{L}_1 + \f{1}{2}(a-ib)\bar{L}_{-1} \label{eq:effectiveH1}
\ee
The Casimir for $H = x^{(0)}L_0 + x^{(1)}L_1 +x^{(-1)}L_{-1} $ is 
\be
C_{SL(2,\bm{R})} = 4 x^{(1)}x^{(-1)} - (x^{(0)})^2.
\ee
Then, for the Hamiltonian (\ref{eq:effectiveH1}) the value of the Casimir becomes
\be
C_{SL(2,\bm{R})} = 4 \f{1}{2}(a-ib)\f{1}{2}(a+ib) - a^2 = b^2 \ge 0 \label{eq:casimir1}
\ee

The Poincare time translation acts on the Poincare coordinate as 
\be
\begin{pmatrix}
t_P \\ x 
\end{pmatrix}
\to 
\begin{pmatrix}
t_P \\ x 
\end{pmatrix}
+\begin{pmatrix}
T_0 \\ 0 
\end{pmatrix}
\ee
and the Rindler boost acts on the Poincare coordinate as 
\be
\begin{pmatrix}
t_P \\ x 
\end{pmatrix}
\to 
\begin{pmatrix}
\cosh h T_1 & \sinh hT_1 \\
\sinh hT_1 & \cosh h T_1
\end{pmatrix}
\begin{pmatrix}
t_P \\ x 
\end{pmatrix}.
\ee
The Floquet one cycle driving is 
\be
\begin{pmatrix}
t_P^{n+1} \\ x ^n 
\end{pmatrix}
\to
\begin{pmatrix}
\cosh h T_1 & \sinh hT_1 \\
\sinh hT_1 & \cosh h T_1
\end{pmatrix}
\begin{pmatrix}
t_P^n \\ x ^n
\end{pmatrix}
+
\begin{pmatrix}
T_0 \\ 0  
\end{pmatrix}
\ee

In terms of $x^{\pm} = x \pm t_P$,the Rindler boost acts on the Poincare coordinate as 
\be
x^+ \mapsto e^{h T_1 }x^+, \qquad x^- \mapsto e^{-h T_1 }x^-.
\ee
The Poincare time translation acts 
\be
x^+ \mapsto x^+ + T_0, \qquad x^- \mapsto x^- - T_0
\ee
In the $SL(2,\mathbb{R})$ notation, the Rindler boost becomes 
\be
x^+ \mapsto \f{e^{\f{hT_1}{2}}x^+ + 0 }{ 0 \times x^+  + e^{-\f{hT_1}{2}}} ,\qquad x^- \mapsto \f{e^{-\f{hT_1}{2}}x^+ + 0 }{ 0 \times x^+  + e^{\f{hT_1}{2}}}.s
\ee
and the Poincare time translation becomes 
\be
x^+ \mapsto \f{x^+ + T_0}{0 \times x^+ + 1}, \qquad  x^- \mapsto \f{x^- - T_0}{0 \times x^- + 1}.
\ee
Therefore, the one cycle is given by 
\be
x^+ \mapsto \begin{pmatrix}e^{\f{hT_1}{2}} & 0 \\ 0 & e^{-\f{hT_1}{2}} \end{pmatrix}  
\begin{pmatrix}  1& T_0 \\ 0 & 1 \end{pmatrix}\cdot x^+ =  \begin{pmatrix}e^{\f{hT_1}{2}}  & e^{\f{hT_1}{2}}T_0 \\ 0 &e^{-\f{hT_1}{2}}  \end{pmatrix} \cdot x^+,
\ee
 and 

\be
x^-\mapsto \begin{pmatrix}e^{-\f{hT_1}{2}} & 0 \\ 0 & e^{\f{hT_1}{2}} \end{pmatrix}  
\begin{pmatrix}  1& -T_0 \\ 0 & 1 \end{pmatrix}\cdot x^- =  \begin{pmatrix}e^{-\f{hT_1}{2}}  & -e^{-\f{hT_1}{2}}T_0 \\ 0 &e^{\f{hT_1}{2}}  \end{pmatrix} \cdot x^-,
\ee
After $n$ cycles, the transformation becomes 
\ba
x^+ &\mapsto \begin{pmatrix} e^{\f{n}{2}hT_1} & e^{-\f{n-2}{2}h T_1}(1 + e^{hT_1} + \cdots e^{(n-1)hT_1})T_0 \\ 0 & e^{-\f{n}{2}hT_1}  \end{pmatrix}\cdot x^+ \notag \\
&= \begin{pmatrix}  e^{\f{n}{2}hT_1} & T_0 (1+ \f{1}{\tanh (\f{1}{2} hT_1)}) \sinh \f{n}{2} hT_1 \\ 0 & e^{-\f{n}{2}hT_1}  \end{pmatrix} \cdot x^+.
\ea
\ba
x^- &\mapsto \begin{pmatrix} e^{-\f{n}{2}hT_1} &- e^{-\f{n}{2}h T_1}(1 + e^{hT_1} + \cdots e^{(n-1)hT_1})T_0 \\ 0 & e^{\f{n}{2}hT_1}  \end{pmatrix}\cdot x^- \notag \\
& = \begin{pmatrix}  e^{-\f{n}{2}hT_1} & -T_0 (-1+ \f{1}{\tanh (\f{1}{2} hT_1)}) \sinh \f{n}{2} hT_1 \\ 0 & e^{\f{n}{2}hT_1}  \end{pmatrix} \cdot x^-.
\ea

\subsubsection{High frequency limit}
We now consider the following high frequency limit 
\be
h T_1 \to  0, \qquad n \to \infty, \qquad n(T_0 + T_1) = t, \qquad  \f{T_1}{T_0+T_1} = \lambda.
\ee
In this limit, the chiral and anti-chiral coordinates are given as 
\ba
x^+(x,t) &=  x e^{h\lambda t} + \f{2 (1-\lambda)}{h \lambda} e^{\f{h}{2}\lambda t}\sinh \f{h\lambda t}{2},  \notag \\
x^-(x,t) &=  x e^{-h\lambda t} - \f{2 (1-\lambda)}{h \lambda} e^{-\f{h}{2}\lambda t}\sinh \f{h\lambda t}{2} . \label{eq:RPfloquetcoord2}
\ea
In the $\lambda \to 0$ limit, this recovers the definition of the holomorphic and anti-holomorphic coordinates $x^{\pm} = x \pm t $. On the other hand, For $\lambda =1$ case this recovers the coordinate transformation to the Rindler coordinate be $x^{\pm} = x e^{\pm h t}$.

The flat metric in these coordinates is 
\be
\text{d} s^2 = - (1 + (-1+h x)\lambda)^2 \text{d}t^2 + \text{d}x^2 .
\ee
The fixed point of the time translation generator is 
\be
1 + (-1+h x_*)\lambda = 0 \qquad \rightarrow \qquad x_* = -\f{1-\lambda}{ h\lambda}.
\ee

\subsubsection{OTOCs for Rindler-Floquet drives}
The Euclidean version of the Rindler time evolution is 
\be
\begin{pmatrix}
\tau_P \\ x 
\end{pmatrix}
\to 
\begin{pmatrix}
\cos h \tau_1 & \sin h \tau_1 \\
-\sin h \tau_1 & \cos h \tau_1
\end{pmatrix}
\begin{pmatrix}
\tau_P \\ x 
\end{pmatrix}.
\ee
Here the Lorentzian time is related to the Euclidean time by $t_P = -\ii \tau_P$ and $T_1 = - \ii \tau_1$.
In the complex coordinate $z = x + \ii \tau_P$, the transformation becomes 
\be
z \to e^{\ii h \tau_1} z.
\ee
The Euclidean version of the Poincare evolution is 
\be
z \to z + \ii \tau_0.
\ee
The Floquet driving in Euclidean signature is then 
\be
z \to \begin{pmatrix}e^{\ii\f{h\tau_1}{2}} & 0 \\ 0 & e^{-\ii\f{h\tau_1}{2}} \end{pmatrix}  
\begin{pmatrix}  1& \ii \tau_0 \\ 0 & 1 \end{pmatrix}\cdot z =  \begin{pmatrix}e^{\ii\f{hT_1}{2}}  & \ii e^{\ii \f{h\tau_1}{2}}T_0 \\ 0 &e^{-\ii\f{h\tau_1}{2}}  \end{pmatrix} \cdot z,
\ee
and 
\be
\bar{z} \to \begin{pmatrix}e^{-\ii\f{h\tau_1}{2}} & 0 \\ 0 & e^{\ii\f{h\tau_1}{2}} \end{pmatrix}  
\begin{pmatrix}  1& -\ii \tau_0 \\ 0 & 1 \end{pmatrix}\cdot \bar{z} =  \begin{pmatrix}e^{-\ii\f{hT_1}{2}}  & -\ii e^{ -\ii \f{h\tau_1}{2}}T_0 \\ 0 &e^{\ii\f{h\tau_1}{2}}  \end{pmatrix} \cdot \bar{z},
\ee
Therefore the coordinate after $n$ cycle driving is 
\ba
z_n & = z_0 e^{\ii h \tau_1}  - \tau_0 \Big(1 - \f{\ii}{\tan \f{h\tau_1}{2}} \Big) e^{\ii \f{h n \tau_1}{2}} \sin \f{n h \tau_1}{2}, \notag \\
\bar{z}_n & = \bar{z}_0 e^{-\ii h \tau_1}  -\tau_0 \Big(1 + \f{\ii}{\tan \f{h\tau_1}{2}} \Big) e^{-\ii \f{hn\tau_1}{2}} \sin \f{n h \tau_1}{2} 
\ea
In the high frequency limit, the coordinate transformation after analytic continuation is
\begin{align}
\label{mapsrindler}
    \tilde{z}(t) = x_0 e^{-h\lambda t}-\frac{2(1-\lambda)}{h\lambda}e^{-h\lambda t/2}\sinh\frac{h\lambda t}{2},\\
\label{mapsrindler2}
\tilde{\bar{z}}(t) = x_0 e^{h\lambda t}+\frac{2(1-\lambda)}{h\lambda}e^{h\lambda t/2}\sinh\frac{h\lambda t}{2},
\end{align}

It is straightforward to understand the flow as $t\rightarrow\infty$ of $\tilde{z}(t)$ and $\tilde{\bar{z}}(t)$:
\begin{equation}
    \lim_{t\rightarrow\infty}\tilde{z}(t) = x_*,\quad
    \lim_{t\rightarrow\infty}\tilde{\bar{z}}(t) = \infty \quad \text{if $x_0>x_*$},\quad
    \lim_{t\rightarrow\infty}\tilde{\bar{z}}(t) = -\infty \quad \text{if $x_0<x_*$},
\end{equation}
i.e., $x_*$ is a stable fixed point for the holomorphic part, and an unstable fixed point for the anti-holomorphic part.
With this in mind, we can now consider that the maps~\eqref{mapsrindler},~\eqref{mapsrindler2} encode the continuous time evolution with the Floquet Hamiltonian $H_{\text{F}}$ in the high-frequency limit. Doing so, the analytic continuation for the cross-ratio is the same as described previously: we simply need to replace $t\rightarrow t+\ii\epsilon_i$, where $\epsilon_i$ will satisfy the correct ordering prescription in order to have the correct analytic continuation. Then, we can compute the cross ratio $\eta$ for the holomorphic part and $\bar{\eta}$ for the anti-holomorphic part. The different scenario for the winding of $\eta$ and $\bar{\eta}$ are summarized in Fig.~\ref{Rindler_floquetdriveOTOC}. The conclusion is that non-trivial OTOC time evolution only happens if $x_1$ and $x_2$, the initial positions of the two fields at time $t=0$, are on the same side of the Floquet horizon $x_*$. 

\begin{figure}[h!]
\begin{center}
	\includegraphics[width=120mm]{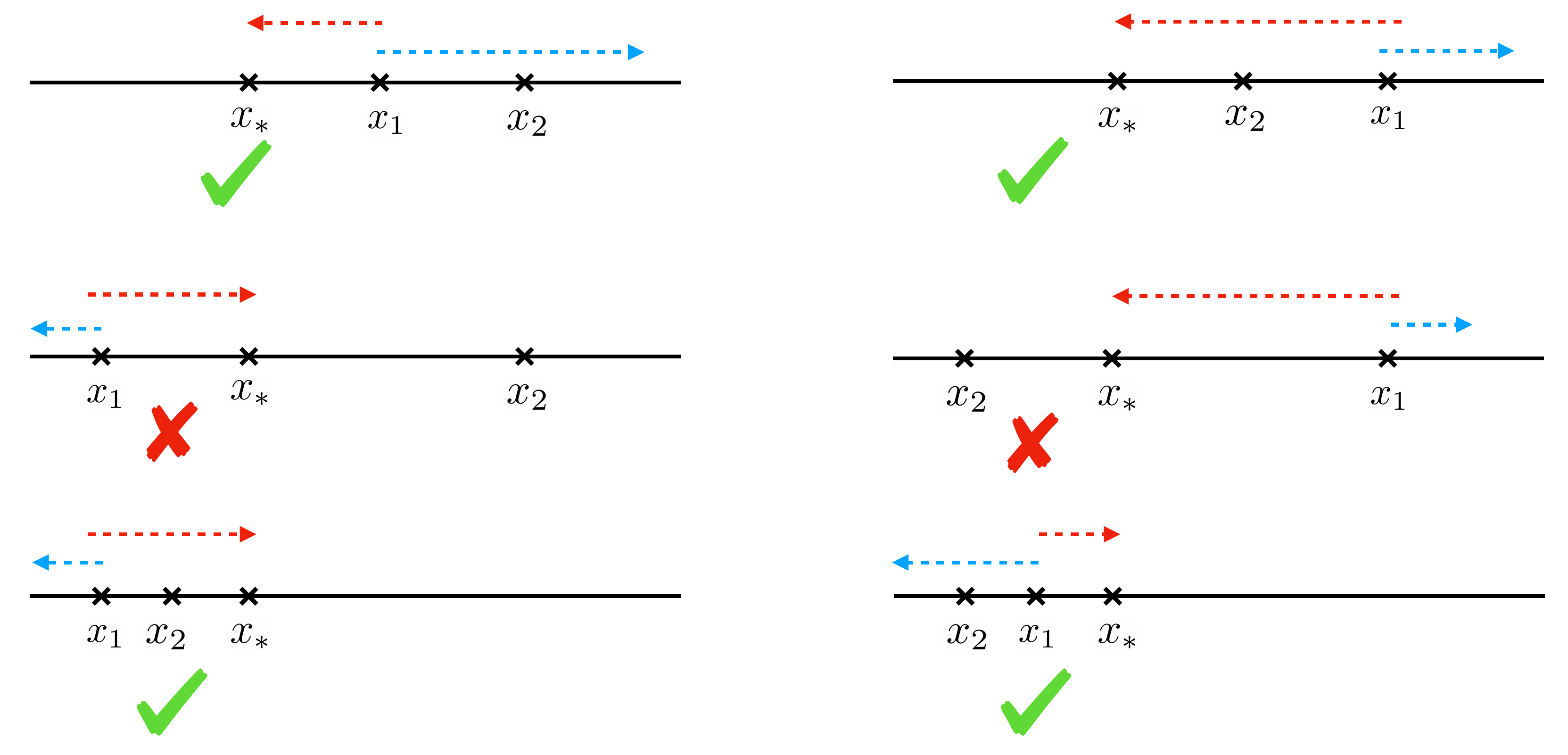}
	\caption{All possible scenarios for the cross-ratio winding depending on the initial positions $x_1$ and $x_2$ of the fields $W$ and $V$ at time $t=0$, relative to the position of the Floquet horizon $x_*$. The red (blue) dashed lines show the flow under $\tilde{z}(t)$ ($\tilde{\bar{z}}(t)$). In order to have a non-trivial OTOC, we need that either $\eta$ or $\bar{\eta}$ goes through the branch cut $[1,\infty)$, which happens only if either the holomorphic or the anti-holomorphic component of the field inserted at $x_1$ at time zero goes through $x_2$ at a finite value of time $t$. The conclusion is that such a non-trivial OTOC only happens with $x_1$ and $x_2$ are on the same side of the Floquet horizon $x_*$. }
\label{Rindler_floquetdriveOTOC}
\end{center}
\end{figure}

Let us now proceed with the computation of the late-time value of the cross-ratio, which determines the time-evolution of the OTOC after the Floquet drive. Essentially, in the large $t$ limit, the cross ratio takes the form
\begin{equation}
\eta \sim \frac{\epsilon_{12}\epsilon_{34}}{\epsilon_{13}\epsilon_{24}}e^{-ht\lambda}\frac{1+\lambda(hx_2-1)}{1+\lambda(hx_1-1)}
\end{equation}
From this value of the cross-ratio at late times we can deduce that the OTOC will decay exponentially with a rate given by $h\lambda$. Furthermore, we can again identify a butterfly velocity, by defining
\begin{equation}
v_*(x_1,x_2) = \frac{1+\lambda(hx_2-1)}{1+\lambda(hx_1-1)}
\end{equation}
We deduce that the Butterfly velocity is precisely given by 
\begin{equation}
v_B(x) =h\lambda x +(1-\lambda).
\end{equation}

\subsection{Stroboscopic backward time evolution}
\label{App:strob_backward_CFT}

In this section we prove the existence of driving parameters $(\widetilde{T}_0,\widetilde{T}_1)$ that satisfy~\eqref{reversalcondition} for a two step drive between two inhomonogeous CFTs. 
We first consider the case of $\mathfrak{sl}(2)$ drives, and consider a two step drive between two Hamiltonians $H_0$ and $H_1$ of the form~\eqref{eq:generalSL2deformation} for durations $T_0$ and $T_1$, with the constraint that $v_0(x)$ and $v_1(x)$ are both strictly positive. In this case choose our driving parameters $(\widetilde{T}_0/L,\widetilde{T}_1/L)$ in such a way that 
\begin{equation}
\label{conditionficedp}
\tilde{\gamma}_1 = \gamma_2, \quad \tilde{\gamma}_2=\gamma_1, \quad \quad \text{stable fixed point} \leftrightarrow \text{unstable fixed point}.
\end{equation}
To prove that this implies~\eqref{reversalcondition}, we simply need to show that the Floquet Hamiltonian of the new driving sequence flips sign $\widetilde{H}_F =-H_{\text{F}}$ as we exchange $\gamma_1\leftrightarrow \gamma_2$.
The change of driving parameters~\eqref{conditionficedp} corresponds to interexchanging the ``sources'' and ``sinks'' of energy and entanglement. From a quasiparticle standpoint, the quasiparticles which were accumulating at the unstable fixed points $x_{*,2}=\frac{L}{2\pi \ii}\log\gamma_2$, $L-x_{*,2} = \frac{L}{2\pi \ii}\log\gamma_2^{*}$ are suddenly repelled as these become stable fixed points, and are attracted by the new unstable fixed points $x_{*,1}=\frac{L}{2\pi \ii}\log\gamma_1$, $L-x_{*,1} = \frac{L}{2\pi \ii}\log\gamma_1^{*}$. While ultimately new horizons will form at these two new positions, there is an intermediate time-scale at which all quasiparticles emitted at time $t=0$ at all spatial positions go back to their initial position, reinitializing the system. 
From a geometric point of view the unit cell of the phase diagram is $(T_0/L,T_1/L)\in[0,1/\mathcal{C}_0]\times [0,1/\mathcal{C}_1]$, with $\mathcal{C}_i$ defined in App.~\ref{app:effectivedescri}. Solving the equation~\eqref{conditionficedp} is equivalent to choosing the new driving parameters $(\widetilde{T}_0/L,\widetilde{T}_1/L)$ to be 
\begin{equation}
\widetilde{T}_0/L =1/\mathcal{C}_0-T_0/L , \quad \widetilde{T}_1/L = 1/\mathcal{C}_1 -T_1/L.
\end{equation}
 This is easily understood by looking at $\gamma_2$ and $\tilde{\gamma}_1$, see Fig.~\ref{fig:exortapt}. For such choice of driving parameters, the condition~\eqref{reversalcondition} is fulfilled as this provides a representation of the inverse Floquet unitary $U_{\text{F}}^{\dagger}$.
\begin{figure}[h!]
\begin{center}
	\includegraphics[width=14cm]{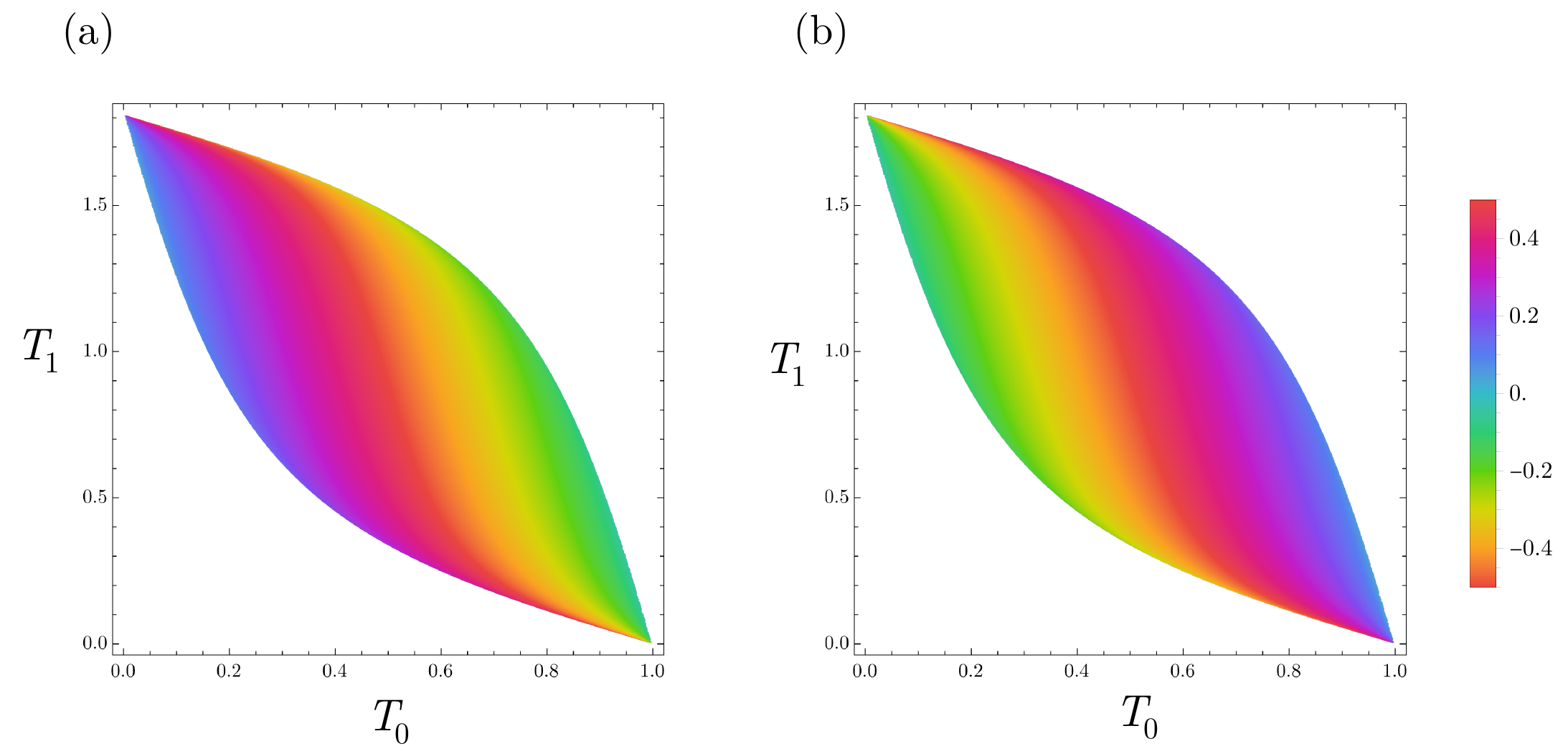}
	\caption{\label{fig:exortapt} (a) Unstable fixed point $\text{arg}(\gamma_2)$ as function of the driving parameters $(T_0,T_1)$ for the first sequence of the drive. Only the heating phase is shown, in which case $|\gamma_{1,2}|=1$. (b) Stable fixed point $\text{arg}(\gamma_1)$ as function of the new driving parameters $(\widetilde{T}_0,\widetilde{T}_1
 )$ for the second sequence of the drive. We show a single unit cell of the phase diagram, and take a drive between $v_0(x)=1$ and $v_1(x)=1-\frac{1}{2}\tanh(0.8)\cos(2\pi x/L)$, such that $\frac{L}{2\pi \ii}\log\gamma_{1,2}\in \left[-\frac{1}{2},\frac{1}{2}\right]$ with periodic boundaries. In both subfigures we normalized $L$ to one.}
\end{center}
\end{figure}

We now design a stroboscopic time-reversal operation for the two-step drive protocol between two general inhomogeneous Hamiltonian with smooth positive deformations, as discussed in App.~\ref{App:OTOCcomputation}. Following the strategy outlined in the $\mathfrak{sl}(2)$ case, we take $\widetilde{T}_0/L$ and $\widetilde{T}_1/L$ to be (assuming that $(T_0/L,T_1/L)\in [0,v_0^{-1}]\times [0,v_1^{-1}]$)
\begin{equation}
\widetilde{T}_0/L=v_0^{-1}-T_0/L, \quad \widetilde{T}_1/L=v_1^{-1}-T_1/L,
\label{concludemeasure}
\end{equation}
such that $(\widetilde{T}_0/L,\widetilde{T}_1/L)\in [0,v_0^{-1}]\times [0,v_1^{-1}]$.
Let us denote the map associated to the new one-cycle diffeomorphism $\tilde{f}_{\pm}(x)$, defined as~\cite{lapierre2020geometric}
\begin{equation}
\label{diffeoreverse}
\tilde{f}_{\pm}(x) = f_0^{-1}(f_0(f_1^{-1}(f_1(x)\mp v_1 \widetilde{T}_1))\mp  v_0\widetilde{T}_0).
\end{equation}
Plugging~\eqref{concludemeasure} into~\eqref{diffeoreverse} and using properties of the circle diffeomorphism $f_{\pm}(x)$, we can readily show that
\begin{equation}
\tilde{f}_{\pm} = f^{-1}_{\pm}(x),
\label{relationdiffeo}
\end{equation}
which directly implies~\eqref{reversalcondition}, as the one-cycle diffeomorphism encodes the stroboscopic time evolution of any primary field.
In other words, the double quench protocol with driving parameters $(T_0/L$, $T_1/L)$ followed by $(\widetilde{T}_0/L$, $\widetilde{T}_1/L)$ leads to a perfect time-reversal of primary fields, just like in the $\mathfrak{sl}(2)$ case. This result shows that the time-reversal procedure~\eqref{concludemeasure} is not simply limited to a finite-dimensional class of spatial deformations, but apply to any smooth deformation of the stress tensor $T_{00}(x)$.
\begin{figure}[h!]
\begin{center}
	\includegraphics[width=15cm]{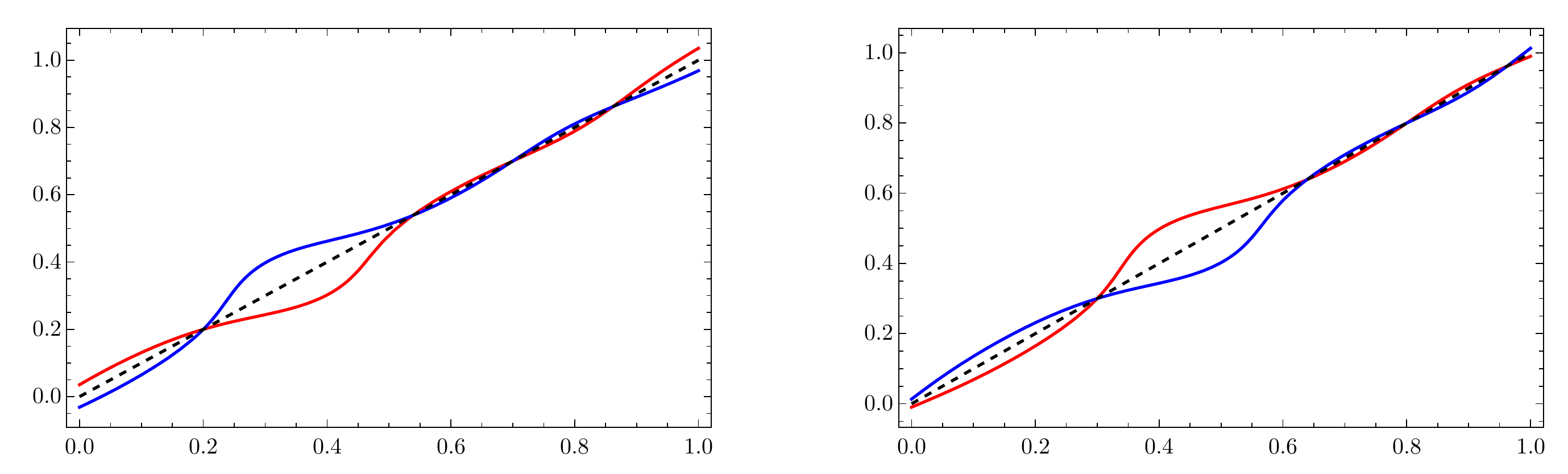}
	\caption{\label{fig:fixedpointsgeneral} Left: One cycle diffeomorphism $f_{+}(x)$ (blue) and its time-reversal partner $\tilde{f}_{+}(x)$ (red). The choice of velocity profile is $v_1(x)=6/(3+\sin4\pi x+\cos2\pi x)$ (which does not lie within the $\mathfrak{sl}(2)$ sector), and $T_0/L=0.1$, $T_1/L=0.45$, $L=1$. We observe four fixed points, two of which are stable and two are unstable. The stability (the sign of $f'_{+}(x_*)-1$ for a fixed point $x_*$) is reversed between the first driving sequence and the second, as a direct consequence of~\eqref{relationdiffeo}. Right: Same, but for the anti-chiral part, $f_{-}(x)$ (blue) and $\tilde{f}_{-}(x)$ (red). }
\end{center}
\end{figure}
In order to gain a geometric understanding of this time-reversal procedure we will make use of the properties of the fixed (or in general, periodic) points associated to the 1-cycle diffeomorphism $f_{\pm}$ in the heating phase. The periodic points of $f_{\pm}$, $\{x_{*,p}^{\mp}\}$ of period $p$, come in pairs of stable and unstable periodic points~\cite{lapierre2020geometric}. The new map $\tilde{f}_{\pm}$ effectively interexchanges each stable and unstable fixed point within each pair, as can be seen from Fig.~\ref{fig:fixedpointsgeneral}, for a choice of velocity profile $v_1(x)$ that involves the full Virasoro algebra. Thus, from a quasiparticle perspective, the source, pumping entangled quasiparticles pairs at each driving cycle, and the sink at which they flow are exchanged, so that each horizon evaporates until the system relaxes back to its initial state.

\subsection{Lattice calculations}

In this section we provide details on the CFT and numerical calculations on a driven free fermion lattice of the fidelity $\text{F}(t)$, as well as energy density and entanglement entropy. The fidelity is in general defined as
\begin{equation}
\text{F}(t) =|\langle\psi_0|U_{\text{F}}^n|\psi_0 \rangle|^2 =|\langle\psi_0| e^{-\ii H_{\text{F}} t}|\psi_0 \rangle|^2,\quad t=nT .
\end{equation}
We consider periodic boundary conditions and choose as initial state the conformal vacuum, $|\psi_0\rangle=|0\rangle$. In this case the time evolution with an $\mathfrak{sl}(2)$ drive is trivial if we choose the generators $\{L_0,L_{\pm1}\}$. However, it leads to a non-trivial evolution of the fidelity for the $n$-fold cover of $\mathfrak{sl}(2)$, i.e., if we choose generators $\{L_0,L_{\pm n}\}$, with $n\geq 2$. Employing two-dimensional representation of the $\mathfrak{sl}(2)$ algebra, one readily finds that in the heating phase
\begin{equation}
\text{F}(t) = \left|\cosh[\pi\Theta_{\text{H}}t]\pm\frac{\gamma_1^*+\gamma_2^*}{\gamma_1^*-\gamma_2^*}\sinh[\pi\Theta_{\text{H}} t]\right|^{-\frac{c}{6}(n^2-1)/n},
\label{fidelity_CFT_drive}
\end{equation}
leading to an exponential decay of the fidelity governed by the Hawking temperature $\Theta_{\text{H}}$ for any $n>1$.

We now consider a 2-step drive between a homogeneous lattice Hamiltonian $H_0$ and a $\mathfrak{sl}(2)$ deformed Hamiltonian $H_1$, both given by~\eqref{lattice_hamitlonians_deformed}, starting from the half-filled Fermi sea as ground state of $H_0$. We denote by $U$ ($V$) the unitary transformation that diagonalizes $H_0$ ($H_1$), and by $\{\epsilon_i^{(0)}\}$ ($\{\epsilon_i^{(1)}\}$) its eigenvalues. In particular, $H_0$ reads in diagonal basis
\begin{equation}
H_0 = \sum_{i=1}^L \epsilon_i^{(0)}\gamma_i^{\dagger}\gamma_i,
\end{equation}
where $\gamma_i = \sum_{j}(U^{\dagger})_{ij}c_j$. Furthermore, we introduce $W=V^{\dagger}U$ for later convenience.
The Loschmidt echo for a single quantum quench with $H_1$ starting from the half-filled Fermi sea thus reads
\begin{equation}
\langle 0|\gamma_{1}...\gamma_{L/2}e^{-\ii H_1t} \gamma_{L/2}^{\dagger}...\gamma_{1}^{\dagger}|0\rangle = \langle 0|(e^{\ii H_1t}\gamma_{1}e^{-\ii H_1t})...(e^{\ii H_1t}\gamma_{L/2}e^{-\ii H_1t})\gamma_{L/2}^{\dagger}...\gamma_{1}^{\dagger}|0\rangle.
\label{loschmidtnumerics11}
\end{equation}
We use that in Heisenberg picture one has
\begin{equation}
\gamma_{k}(t) = e^{\ii H_1t}\gamma_{k}e^{-\ii H_1t}=\sum_{l=1}^{L}(W^{\dagger}\cdot W^{(t),1})_{kl}\gamma_{l},
\label{loschmidtnumerics22}
\end{equation}
where we defined $W^{(t),1}_{il}=e^{-\ii \epsilon^{(1)}_i t}W_{il}$. Inserting~\eqref{loschmidtnumerics22} into~\eqref{loschmidtnumerics11} and using Wick's theorem, we conclude that
\begin{equation}
\text{F}(t)  =\left| \text{det}_{1\leq\ j, l \leq L/2}(W^{\dagger}\cdot W^{(t),1} )_{jl}\right|^2.
\end{equation}
The above derivation readily generalizes to an $n$-cycle of the Floquet drive between $H_0$ and $H_1$, leading to the general formula
\begin{equation}
\text{F}(t)=\left|\text{det}_{1\leq\ j, l \leq L/2} \left(W^{\dagger}\cdot W^{(T_1),1}\cdot\text{diag}(e^{-\ii \epsilon^{(0)}_1T_0},...,e^{-\ii \epsilon^{(0)}_LT_0})\right)^n_{jl}\right|^2,
\label{fidelity_lattice_drive}
\end{equation}
The agreement between~\eqref{fidelity_CFT_drive} (with central charge $c=1$) and~\eqref{fidelity_lattice_drive} is illustrated in the case of a two-step Floquet drive in both heating and nonheating phases in Fig.~\ref{fig:fidelity_periodic_drive}.
\begin{figure}[h!]
\begin{center}
	\includegraphics[width=15cm]{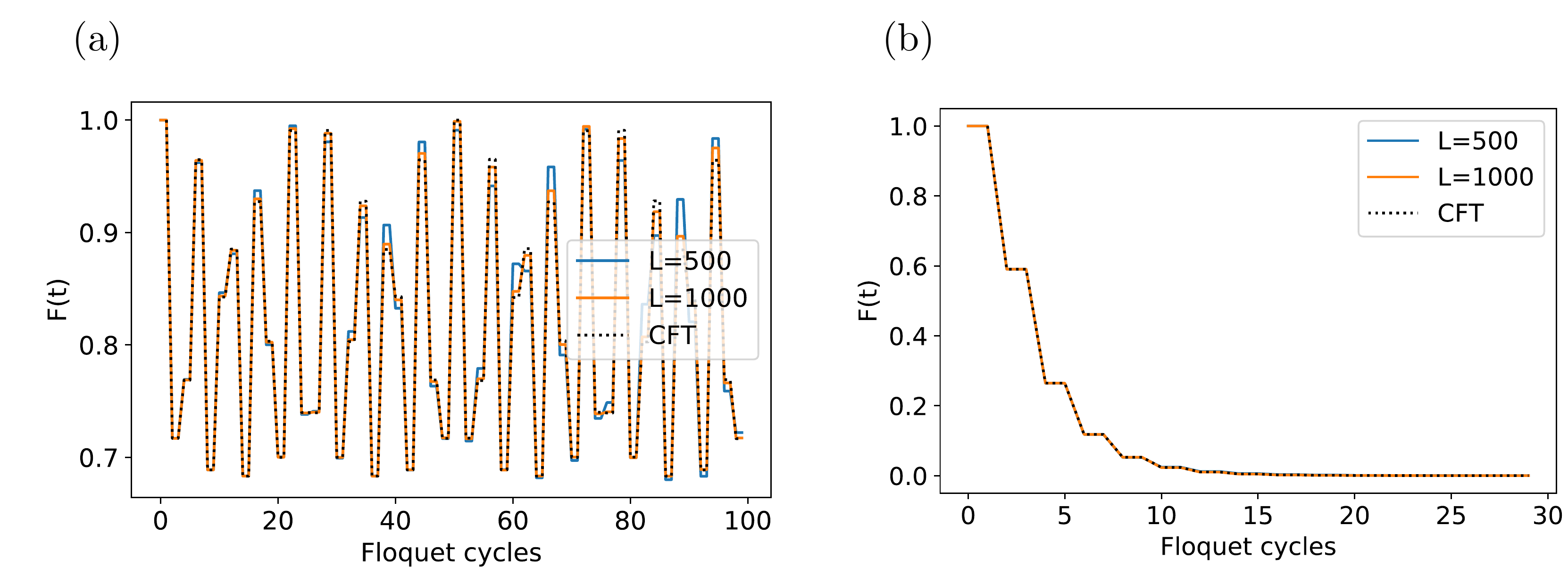}
	\caption{\label{fig:fidelity_periodic_drive} Stroboscopic fidelity evolution $\text{F}(t)$ for a 2-step drive between $H_0$ and $H_1$ with a deformation $v_1(x)=2\sin^2\left(\frac{3\pi x}{L}\right)$, in (a) the nonheating phase and (b) the heating phase. The agreement between the CFT~\eqref{fidelity_CFT_drive} predictions with $c=1$ and the lattice formula~\eqref{fidelity_lattice_drive} is manifest.}
\end{center}
\end{figure}
On the other hand, the energy density and entanglement entropy are both 
readily obtained on the lattice from the equal-time correlation function~\cite{Peschel_2009}
\begin{equation}
C_{ij}(t) = \langle c^{\dagger}_i(t)c_j(t) \rangle,
\end{equation}
which for the 2-step drive reads~\cite{floquetcftwen}
\begin{equation}
C_{ij}(t) = \sum_{k=0}^{L/2}\left(U\cdot\left[W^{(T_0),0}\cdot W^{(T_1),1}\right]^n\right)_{jk}  \left(U\cdot\left[W^{(T_0),0}\cdot W^{(T_1),1}\right]^n\right)^{\dagger}_{ki} .
\end{equation}

We now provide numerical details on the fidelity of the time-reversal procedure described in App.~\ref{App:strob_backward_CFT}. In this case, we consider a generalization of~\eqref{fidelity_lattice_drive} to the case of a ``double-drive'' protocol,
\begin{equation}
\text{F}(t)=\left|\text{det}_{1\leq\ j, l \leq L/2} \left(\left[W^{\dagger}\cdot W^{(T_1),1}\cdot\text{diag}(e^{-\ii \epsilon^{(0)}_1T_0},...,e^{-\ii \epsilon^{(0)}_LT_0})\right]^n\left[W^{\dagger}\cdot W^{(\widetilde{T}_1),1}\cdot\text{diag}(e^{-\ii \epsilon^{(0)}_1\widetilde{T}_0},...,e^{-\ii \epsilon^{(0)}_L\widetilde{T}_0})\right]^n\right)_{jl}\right|^2,
\label{fidelity_lattice_drive2}
\end{equation}
which is the fidelity of the time-reversal procedure after driving the system for $n$-cycles with initial parameters $(T_0,T_1)$. While the time-reversal procedure is exact for CFTs, as shown in App.~\ref{App:strob_backward_CFT}, finite-size lattice effects irremediably lead to a breakdown of the fidelity for a large number of driving cycles. The precise number of such cycles depends non-trivially on the length of the chain, as well as on the driving parameters $(\frac{T_0}{L},\frac{T_1}{L})$ that determine the heating rate. In fact, the higher the heating rate, the faster the deviation between lattice and CFT will be evident as the lattice system leaves its low-energy sector.
\begin{figure}[h!]
\begin{center}
	\includegraphics[width=16cm]{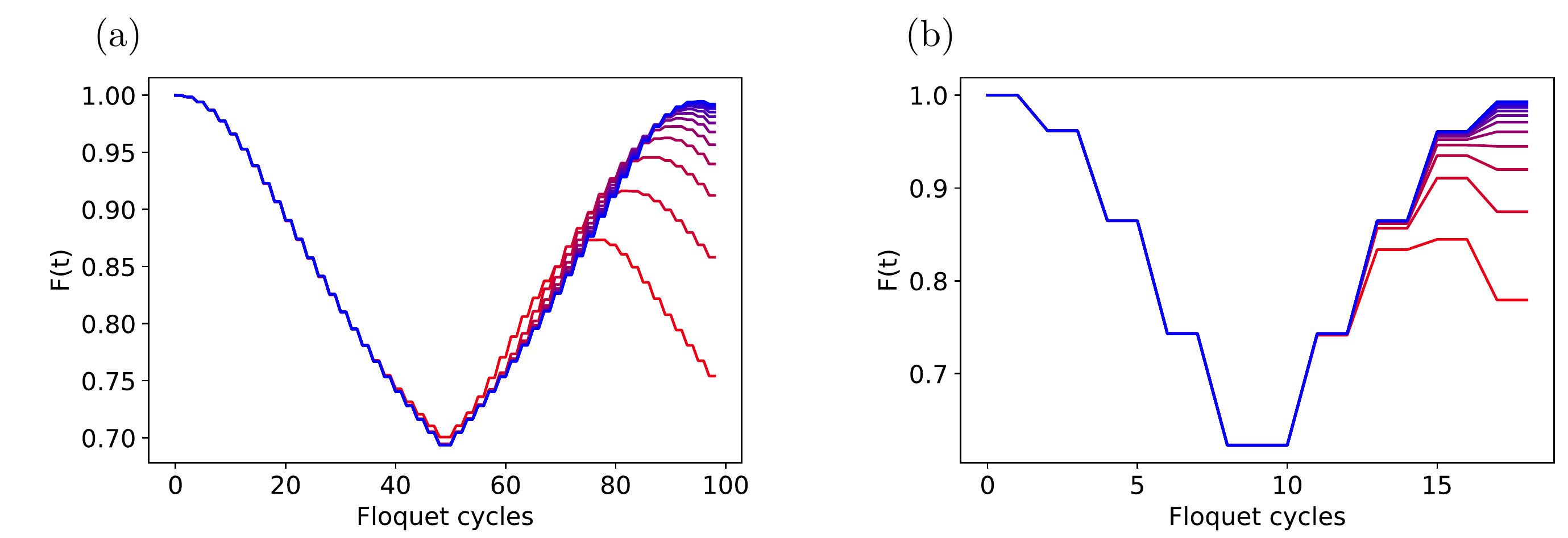}
	\caption{\label{fig:convergence_loschmidt} Fidelity $F(t)$ after the time-reversal protocol on the lattice for system sizes $L=100, 200, ..., 1200$ (red to blue), for $v_1(x) = 1-\tanh(0.4)\cos(4\pi x/L)$, and driving parameters (a) $\frac{T_0}{2L}=0.96$, and $\frac{T_1}{2L\cosh(0.4)} = 0.06$, (b) $\frac{T_0}{2L}=0.5$, and $\frac{T_1}{2L\cosh(0.4)} = 0.5$.}
\end{center}
\end{figure}
This is illustrated on Fig.~\ref{fig:convergence_loschmidt}, where by approaching the thermodynamic limit the fidelity $F(t)$ converges towards a maximal value. The number of cycles for which the procedure is faithful is typically shown on Fig.~\ref{fig:convergence_loschmidt}(a) for a low value of the heating rate and on Fig.~\ref{fig:convergence_loschmidt}(b) for the maximal value of the heating rate for the given driving protocol.

\end{widetext}
\end{document}